\documentclass[useAMS,usenatbib]{mn2e}
\usepackage{graphicx}
\usepackage{lmodern}
\usepackage{slantsc}
\pdfoutput=1

\title[Evolution of Prolate Clouds at H\hspace{1pt}{\textsl{\textsc{ii}}} Boundaries]{Evolution of Prolate Molecular Clouds at \HII\ Boundaries: I. Formation of fragment-core structures}
\author[T. M. Kinnear, J. Miao, G. J. White and S. Goodwin]{T. M. Kinnear$^{1}$\thanks{E-mail:
tk218@kent.ac.uk}, J. Miao$^{1}$, G. J. White$^{2,3}$ and S. Goodwin$^{4}$ \\
$^{1}$ Centre for Astrophysics and Planetary Science, School of Physical Sciences, University of Kent, Canterbury, CT2 7NH, England \\ 
$^{2}$ Department of Physics and Astronomy, The Open University, Milton Keynes MK7 6AA , England \\
$^{3}$ Space Science and Technology Department,CCLRC Rutherford Appleton Laboratory, Oxfordshire OX11 0QX, England \\ 
$^{4}$ Department of Physics and Astronomy, University of Sheffield, Sheffield, S3 7RH, UK
}

\newcommand{\msun}{M$_{\odot}$}
\newcommand{\htwo}{H$_2$}
\newcommand{\htwoden}{cm${}^{-3}$}
\newcommand{\HII}{H\hspace{1pt}{\sc ii}}
\newcommand{\CI}{C\hspace{1pt}{\sc i}}
\newcommand{\CII}{C\hspace{1pt}{\sc ii}}
\newcommand{\OI}{O\hspace{1pt}{\sc i}}
\newcommand{\OII}{O\hspace{1pt}{\sc ii}}

\begin{document}

\date{Accepted 20?? Month ??. Received 20?? Month ??; in original form 20?? Month ??}

\pagerange{\pageref{firstpage}--\pageref{lastpage}} \pubyear{20??}

\maketitle

\label{firstpage}

\begin{abstract}
The evolution of a prolate cloud at an \HII\ boundary is investigated using Smoothed Particle Hydrodynamics (SPH). The prolate molecular clouds in our investigation are set with their semi-major axis perpendicular to the radiative direction of a plane parallel ionising Extreme Ultraviolet (EUV) flux.

Simulations on three high mass prolate clouds reveal that EUV radiation can trigger distinctive high density core formation embedded in a final linear structure. This contrasts with results of the previous work in which only an isotropic Far Ultraviolet (FUV) interstellar background flux was applied.

A systematic investigation on a group of prolate clouds of equal mass but different initial densities and geometric shapes finds that the distribution of the cores over the final linear structure changes with the initial conditions of the prolate cloud and the strength of the EUV radiation flux. These highly condensed cores may either scatter over the full length of the final linear structure or form two groups of high density cores at two foci, depending on the value of the ionising radiation penetration depth $d_{\mathrm{EUV}}$, the ratio of the physical ionising radiation penetration depth to the minor axis of the cloud. Data anlysis on the total mass of the high density cores and the core formation time finds that the potential for EUV radiation triggered star formation efficiency is higher in prolate clouds with shallow ionisation penetration depth and intermediate major to minor axial ratio, for the physical environments investigated.

Finally, it is suggested that the various fragment-core structures observed at \HII\ boundaries may result from the interaction between ionising radiation and pre-existing prolate clouds of different initial geometrical and physical conditions. 

\end{abstract}


\section{Introduction}
Newly formed massive stars emit intense UV radiation onto the surfaces of surrounding molecular clouds, ionizing and heating gas on their star-facing surfaces. The ionisation heating ejects the ionised gas from the cloud to create a hot and diffuse \HII\ region, whilst at the same time drives a compressive wave toward the interior of the cloud to form condensed core(s), in which new star(s) could form. This is the so-called Radiative Driven Implosion (RDI) process \citep{Bertoldi1989-1}. Additionally,  emission from the recombination of electrons with ions creates a bright rim at the edge of the molecular cloud on its star-facing side. The resultant cloud structure containing a bright rim and condensed core is termed a Bright Rimmed Cloud (BRC), which has interested astronomers over the last two decades and their study has been an important observational step in the development of models of triggered star formation \citep{ElmegreenLada1977-1,McKeeHollenbach1980-1,SandfordEtAl1982-1}. 

The majority of observed BRCs can be categorised into three types according to their morphologies, types A, B and C in an order of increased curvature of their bright rims \citep{SugitaniEtAl1991-1, SugitaniOgura1994-1, SugitaniEtAl1995-1}. Recent observations have revealed more intriguing structural features of BRCs, such as fragment-core structures perpendicular to the radiation flux direction \citep{ChauhanEtAl2011-2}; cometary type C structures not aligned to the direction of the incident radiation \citep{OguraSugitani1998-1,MorganEtAl2004-1,KarrEtAl2005-1,UrquhartEtAl2006-1, FukudaEtAl2013-1}; RDI triggered multi-star formation in BRCs \citep{ChoudhuryEtAl2010-1,ChauhanEtAl2011-1} and symmetrical BRC structures sandwiched between two \HII\ regions \citep{CohenEtAl2003-1,OjhaEtAl2011-1}.

Based on the RDI mechanism, current theoretical investigations have successfully revealed a possible physical process  for the formation of a BRCs having symmetrical morphologies \citep{Bertoldi1989-1, LeflochLazareff1994-1, LeflochLazareff1995-1,KesselBurkert2000-1,KesselBurkert2003-1,EsquivelRaga2007-1, MiaoEtAl2006-1,MiaoEtAl2009-1, Gritschneder2009-1,BisbasEtAl2011-1,HaworthHarries2012-1}. Although \citet{MiaoEtAl2010-1} have studied the possibility for the formation of the IC59 structure (type M) \citep{KarrEtAl2005-1}, little attention has been paid to explore the RDI triggered star formation process in   asymmetrical BRCs.

Most of the current RDI models adopt a spherical molecular cloud as the initial condition in simulations. However, recent observations on a large sample of isolated molecular cores have revealed that spherically symmetric molecular cloud cores are the exception rather than the rule \citep{JonesEtAl2001-1,MyersEtAl1991-1,CurryStahler2001-1,RathborneEtAl2009-1}. Theoretical investigations have also found physical mechanisms which result in formation of prolate clouds in general astrophysical environments \citep{Tassis2007-1, Boss2009-1, CaiTaam2010-1}. Shown in Figure 1 is the distribution of molecular clumps over the ratio $\gamma$ of the semi-major ($a$) to semi-minor ($b$) axis,  resulted from a Galactic Ring survey of 6,124 objects, which gives a mean axial ratio $\gamma = \frac{a}{b} = 1.6$  \citep{RathborneEtAl2009-1}. It is worth noting that the data from which these ratios were calculated was determined using the FWHM of two axes in the observation. As such, clumps which are elongated but have their semi-major axis intercepting the observational plane by an angle are represented with a lower $\gamma$ value  than their actual ones. It can be expected that the `true' ratios of these objects will be shifted to higher $\gamma$ value range.

\begin{figure}
\center
\includegraphics[width=0.47\textwidth]{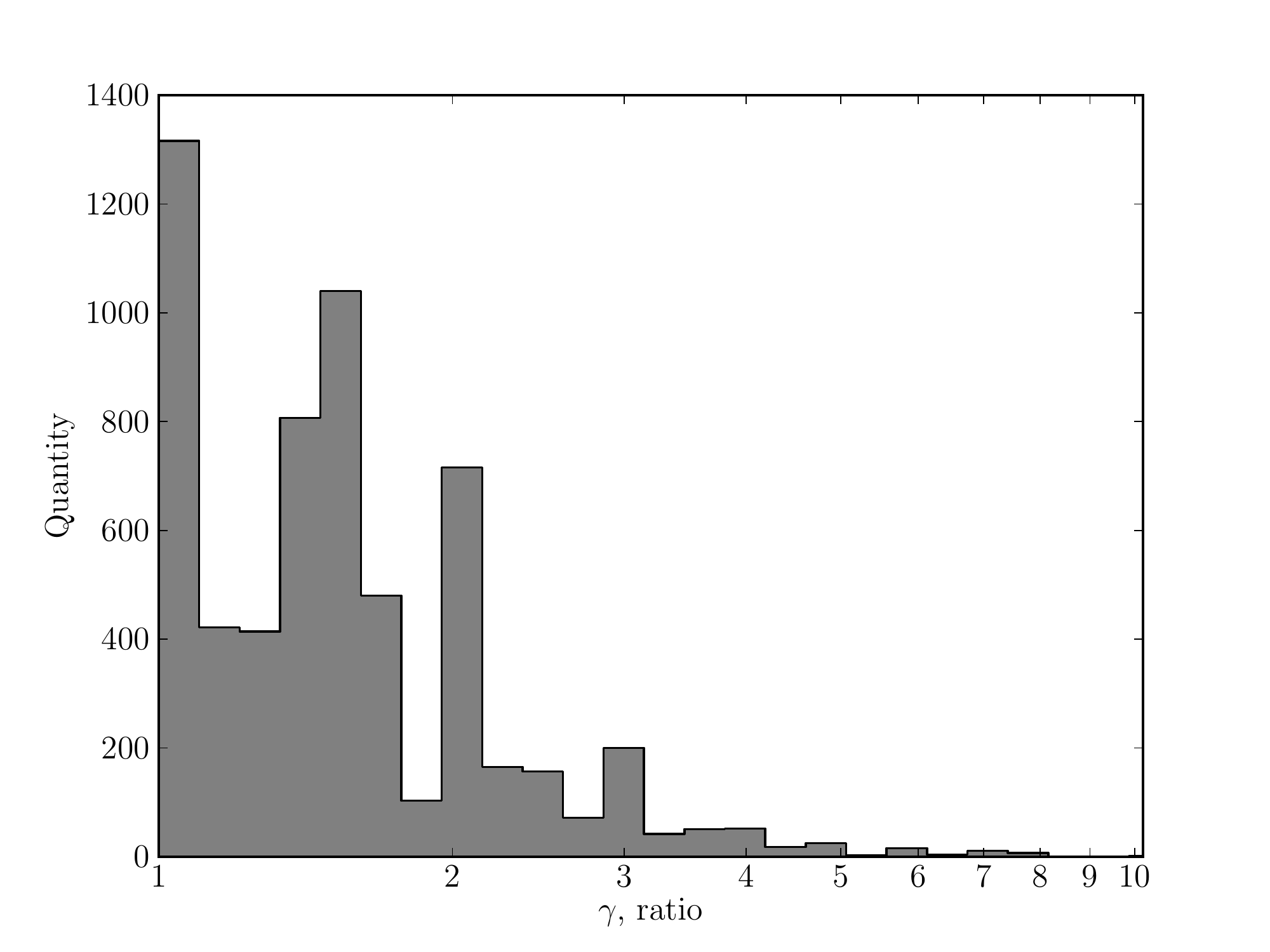}
\caption{The distribution of molecular clumps over the semi-major to minor axial ratios $\gamma$. Statistics based on the data in a survey for 6,124 single molecular clumps \citep{RathborneEtAl2009-1}. }
\label{ratiodistribution}
\end{figure}

Therefore, the assumption of an initially spherical molecular cloud in theoretical modelling may be too simplistic for a complete view of the diverse structures found at \HII\ boundaries. Although some previous work has investigated the collapse of a prolate cloud subject to an isotropic FUV radiation field \citep{NelsonLanger1997-1}, the dynamical evolution of a prolate cloud at an \HII\ boundary has not yet been investigated. Therefore we have attempted to investigate this scenario with prolate molecular clouds of various initial geometries and physical conditions. Our objective is to explore possible physical mechanisms for a variety of structures found at \HII\ boundaries but not yet well understood.

 In this paper, we focus on the investigation of the evolution of a prolate cloud at an \HII\ boundary with its semi-major axis perpendicular to the ionisating radiation flux. In Section \ref{codesection} we  briefly describe the numerical codes used along with data processing, as well as the initial conditions of the prolate clouds adopted in our simulations. Our simulation results and discussions are presented in Section \ref{resultsetc} and the conclusions are discussed in Section \ref{conclusion}. Table \ref{overall} describes all of the test series used in the paper, along with their purpose and the section(s) in which they are discussed.

\begin{table*}
\begin{minipage}{1.0\textwidth}
\centering
\begin{tabular}{lllllr@{}c@{}llp{2.3in}}
\cline{1-10}
Name & Mass & Density & Ratio & EUV Flux & \multicolumn{3}{l}{Varied Parameter} & Section & Purpose\\
 & (\msun) & (\htwoden) & & (cm$^{-2}$ s$^{-1}$) &  &\\
A, B, C & $M$ & 100 & 2 & $10^9$ & $100 \leq$&$M$&$\leq 200$ & \ref{NelsonLangerSection} & Observation of the evolution of high $d_{\mathrm{euv}}$ clouds of varied mass. \\
D1-3 & 200 & $n$ & 2 & $10^9$ & $100 \leq$&$n$&$\leq 1,200$ & \ref{Dtests} & Observation of high mass clouds with varied density. \\
E1-3 & 200 & 100 & 2 & $F$ & $10^7 \leq$&$F$&$\leq 8\times10^9$ & \ref{Etests} & Observation of high mass clouds with varied incident flux. \\
G1(1-19) & 30 & 600 & $\gamma$ & $10^9$ & $1 \leq$&$\gamma$&$\leq 8$ & \ref{G1} & Observation of low mass clouds at medium initial density across varied ratios. \\
G2(1-19) & 30 & 1,200 & $\gamma$ & $10^9$ & $1 \leq$&$\gamma$&$\leq 8$ & \ref{G2} & Observation of low mass clouds at high initial density across varied ratios.\\
G0 & 30 & 100 & 2 & $10^9$ &  & & & \ref{G0} & Observation of extending particular ratios to a low initial density (other ratios were produced, but only $\gamma = 2$ is presented). \\
\cline{1-10}
\end{tabular}
\caption{Summary of the parameters of all tests examined. After the name of the test series, the next four columns (Mass, Density, Ratio \& EUV Flux) are the defining parameters of each simulation set. For all series, three of the four are fixed values, and the remaining parameter is represented by a variable, whose range is described in the Varied Parameter column. The final column indicates the main Section/Subsection in which the test series is discussed.}
\end{minipage}
\label{overall}
\end{table*}

\section{The Code and Initial Conditions}
\label{codesection}

\subsection{The code}
All of the simulations presented in this paper were performed using an extended Smoothed Particle Hydrodynamics (SPH) code II, which is based on the SPH code I by \citet{NelsonLanger1997-1}. The latter  was used to investigate the evolution of a molecular cloud in an isotropic interstellar background FUV radiation field. Code I was extended by including EUV radiation transferring into a molecular cloud and the consequent physical processes. Therefore, the recently refined code II contains the following components: i) SPH solvers for the full set of standard hydrodynamic equations (including energy evolution equation); ii) ray-tracing solver for the radiation transferring equations, which is based on the method of \citet{KesselBurkert2000-1}; iii) a numerical solver for a set of chemical reaction differential equations, which evolves the fractional abundances of the chemical species: CO, \CI, \CII, HCO$^+$, O, He$^+$, OH$_x$, CH$_x$, H$_3^+$, M, M$^+$ and free electrons \citep{NelsonLanger1997-1}. Further details of  code II can be found in \citet{MiaoEtAl2006-1}. In the following, we present a brief summary of its main features.    

In the hydrodynamic equation solver, each SPH particle is given an adaptive smoothing length $h$, therefore additional $\nabla h$ terms are included in the equations of motion in order to satisfy conservation requirements \citep{NelsonLanger1994-1}. The value of a function at each particle is calculated by the average of that of $N_{\mathrm{neigh}}=45$ neighbouring particles, weighted by the standard M4 cubic spline kernel function. The equation of state $P=c_v(\gamma -1) \rho T$ is used, where $\rho$ is the gas density,  $T$ is the temperature, $\gamma$ is the ratio of specific heats and $c_v$ the fixed volume specific heat capacity of the gas. The temperature of each particle $T$ is determined by solving the energy conservation equation in the standard hydrodynamic equations, rather than calculated from an assumed function of gas density or ionisation fraction as commonly used  in other existing ionisation codes \citep{LeflochLazareff1994-1, KesselBurkert2000-1, Gritschneder2009-1, BisbasEtAl2011-1}. Following similar reasoning as \citet{BisbasEtAl2011-1}, we take $\gamma = 5/3$. The temperature profile at an \HII\ boundary is very distinctive, with a sharp boundary between ionised atomic gas ($\ge 10^{4}$ K) and neutral gas ($\le 200$ K).  In the latter the rotational degrees of freedom of H$_2$ are only weakly excited, so we can still assume that $\gamma \sim 5/3$ even for H$_{\mathrm{2}}$.     

In the energy conservation equation, the heating rate function is dominated by the term for the hydrogen ionisation heating produced by EUV radiation ($h\nu > 13.6$ eV) from a nearby star and the photoelectric ejection of electrons from dust grains caused by the FUV radiation (6.5 $ < h\nu < 13.6$ eV). The former process is much more effective in heating the gas than the latter, i.e., by two orders of magnitude. In the ionised gas regions the cooling rate function is mainly contributed to by recombination of the electrons with ions and the collisional excitation of \OII\ lines; in the cooler, unionised regions, it is dominated by CO, \CI, \CII\ and \OI\ line emissions.

\subsection{Initial and boundary conditions}

All of the molecular clouds in our simulations start with a uniform density, which is rendered by a glass-like  distribution of SPH particles created using GADGET-2 \citep{Springel2005-1}. Compared with a uniform random distribution, a glass-like distribution has a substantially lower noise in the resulting density distribution. This is of particular benefit in circumstances where small variations are likely to be amplified in the resulting evolution. The number of SPH particles for each molecular cloud is decided according to the mass resolution required by the convergence test of the code II, $10^{-3}$ \msun\ per SPH particle. A zero initial velocity field is set for all of the molecular clouds in the simulations.

\begin{figure}
\center
\includegraphics[width=0.47\textwidth]{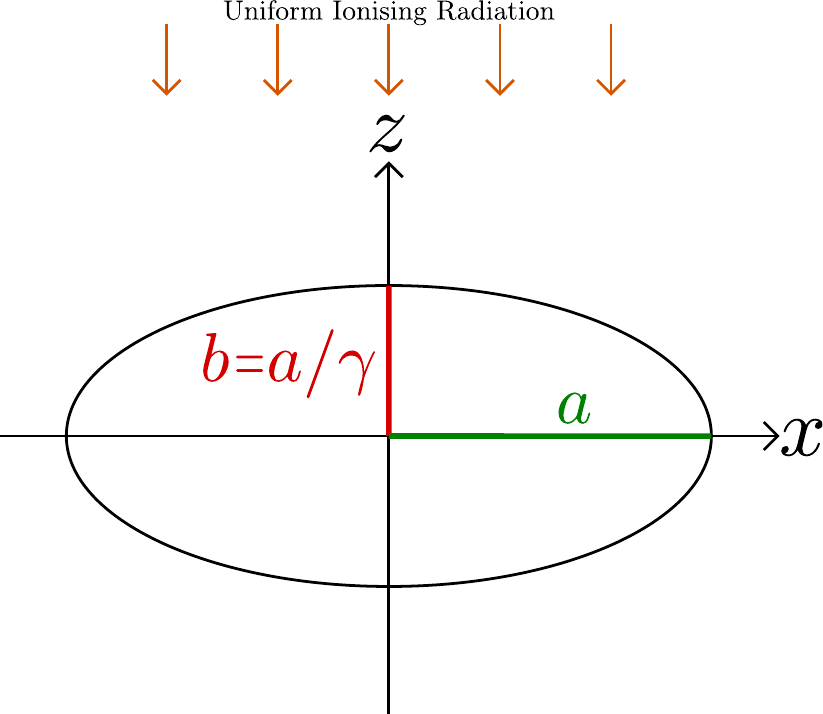}
\caption{The projected two dimensional diagram (onto $xz$ plane) of the initial geometry of a prolate cloud and the configuration of the ionising (EUV) radiation flux from nearby star(s). The isotropic interstellar background FUV is not drawn in the diagram but considered in simulations.}
\label{illustration}
\end{figure}

We investigate the dynamic evolution of a prolate cloud with its semi-major axis perpendicular to the incident direction of EUV radiation as shown in Figure \ref{illustration}. Rather than specifying the initial geometry of the prolate cloud by semi-major and semi-minor axis ($a$, $b$), we use ($a$, $\gamma = \frac{a}{b}$ ) as the pair of initial geometrical parameters for the cloud. The objective of this investigation is to observe the EUV radiation triggered collapse of a prolate cloud, therefore we set the initial geometric parameters of a prolate cloud of mass $M$ in such a way that it would be stable without a radiation field. The Jeans criteria (in terms of Jeans number $J$) for an isolated prolate cloud to be stable against its own gravity can be expressed as \citep{Bastien1983-1},

\begin{equation}
 J = \frac{\pi \; G \; \rho\; \mu\; b^2}{15\; e\; R_{\mathrm{g}}\; T} \ln{\left(\frac{1 + e }{1 - e }\right)} \le 1  
\label{jeansnumber}
\end{equation}
where $\rho$, $b$,  $T$ and $\mu$ are the mass density, 
the minor axis, the initial temperature and the mean molecular mass of the prolate cloud respectively, 
$G$ and $R_{\mathrm{g}}$ the Gravitational Constant and Specific Gas Constant, 
and with the eccentricity $e =  \sqrt{1 - \frac{b^2}{a^2} }= \frac{\sqrt{\gamma^2 - 1}}{\gamma}$.  

Substituting $\rho = \left(3 M\right) / \left(4 \pi a b^2\right) $ into Equation \ref{jeansnumber},  
we get the condition for the major axis $a$ of an isolated non-collapsing prolate cloud       
\begin{eqnarray}\nonumber
a \ge a_{\mathrm{crit}} & = & \frac{\mu\; G\; M }{20\; R_{\mathrm{g}}\; T\; e} \ln \left(\frac{1 + e}{1 - e}\right)\\
&=& 0.052 \; \frac{M^*\; \gamma}{T\; \sqrt{\gamma^2 - 1}} \ln \left(\frac{\gamma + \sqrt{\gamma^2-1}}{\gamma - \sqrt{\gamma^2 - 1}}\right) 
\label{majoraxis}
\end{eqnarray}
where $M^*$ is the mass of the prolate cloud in units of solar masses, and $a$ and $a_{\mathrm{crit}}$ have units of Parsecs.
For a given molecular cloud of mass $M^*$, and initial temperature $T$ and $\gamma$, a minimum value $a$ can be estimated, the major axis of an initially gravitationally stable cloud should satisfy $a > a_{\mathrm{crit}}$.

All of the prolate clouds investigated were subject to an isotropic interstellar background FUV radiation of one Habing unit \citep{Habing1968-1} and an ionising EUV radiation with a flux of $ 10^{9}$ cm$^{-2}$ s$^{-1}$ (typical of the boundary of an \HII\ region) directed parallel to the $z$-axis (along the negative $z$ direction) as illustrated in Figure \ref{illustration}, in which the isotropic interstellar background FUV radiation is not shown, although it is included in our simulations. The boundary condition takes the form of a spherical outflow boundary, with a weak boundary pressure.

  
\subsection{Core finding program}
\label{corefinder}
Because of the occurrence of fragmentation in the evolution of the prolate clouds in our simulations, the number and locations of condensed cores will provide useful information on the potential sites for EUV radiation triggered star formation. The physical properties of cores formed is derived by using a corefinding code developed to recursively `grow' a candidate core outwards from a high density particle, connecting in a tree-like structure to nearby particles of lower density. 

 A `core' in this context is defined as a region surrounding a local density maxima with a peak  \htwo\ number density greater than $10^{6}$ \htwoden. This results in selection of `cores' with a wide range of peak densities, from just over $10^{6}$ \htwoden\ up to the code's effective limit of $\approx 10^{13}$ \htwoden. The occurrence of the upper limit on the number density is because
the Courant-Friedrichs-Lewy (CFL) condition time step used in the code becomes extremely small when a high density of $10^{13}$ \htwoden\ is approached. No sink-particle implementation is implemented in the code, which makes the simulation almost cease to evolve much further after the formation of the first few high density cores. We use this as the definition of the `end' of the simulation, wherever subsequently referred to. 
Therefore, our main interest is to explore the effects of the initial conditions of a prolate cloud on its dynamical evolution up to the first batch of proto-star seed formation.
In the following we present the main frame work of the corefinding code. 

To begin with, all particles below a density threshold ($n < 10^6$ \htwoden) are discounted. Following this initial filter of particles, the cores are determined in the following procedures:
\begin{enumerate}
\item The code generates nearest neighbour lists for every particle. A set of 45 neighbours, the same as $N_{\mathrm{neigh}}$ for the SPH code, will be located and selected for the use of the code presented here.

\item The particle with maximum density is selected and acts as the seed for the first core.
\label{procstartshere}

\item The code then searches outwards to select all of the nearest neighbour particles which have a density lower than the seed particle. From each of these neighbours, the selection process then attempts to search further outward for any lower density neighbours which have not already been selected. Each particle is added to a list for the current seed as they are selected. Two exceptions exist which permit selection of a particle of a higher density than that of the current particle. The first is an `over-density' margin which was set as 1\% of the current density, to select individual spuriously over-dense particles. The second is that any connected particles will be automatically selected, regardless of relative number density, if they are above the jeans density limit described by \citet{BateBurkert1997-1}. This is the density above which artificial fragmentation is expected to occur, and the selection process ensures that local density maxima separated by greater than this density are jointly selected as a single core.

\item The selection process continues until no particles remain which are: (a) lower number density than the last seed particle (plus the two exceptions); (b) not already selected by the current seed.

\item This procedure is then repeated from step \ref{procstartshere}. This time the new seed particle is selected as being the next maximum density particle which has {\bf not} already been selected by a previous descent along nearest neighbour branches. This is done until no particles remain which can be selected, having been ruled out by one or more of the previously described criteria. Note that particles already selected and labelled by one seed may also be selected and labelled by another seed. Each particle builds up a list of which seeds have selected it.

\item Through this process, particles are selected in groups growing out from all localised density maxima. Following the selection of all possible candidate groups, mean properties for each group are determined (position, density, i.e., collective properties of any attribute possessed by the component particles).

\item An additional point of note regards particles which were selected and labelled from multiple seeds. In the current implementation, the properties of any particle which is part of more than one seed descent is equally weighted between those seeds. A more comprehensive process for deciding `ownership' of each particle will be implemented in the near future. 

\end{enumerate}

The method presented here may provide an advantage in determining non-spherical or highly asymmetrical cores, for which a radial selection or search may not be sufficient. It additionally permits determination of structure shapes which are of highly irregular geometries. These include filament and clump features for simulations involving larger scale, clumpier structures than those dealt with in this paper, for the approximate shape, size and extent of each core can be determined in the code.

\subsection{The EUV flux penetration parameter}
The role of the intensive ionising radiation flux on the evolution of molecular cloud is manifested in two important ways. As stated in the RDI model, an ionising radiation induced shock compresses the neutral and cool gas  in a molecular cloud into condensed cores which may collapse to form stars under its enhanced self-gravity. At the same time, ionising radiation induced photo-evaporation erodes gas material from the surface of the cloud, which weakens the potential for star formation. Whether a pre-existing cloud could be triggered to form stars or totally photo-evaporated depends on the two competing effects of an EUV radiation field.

To classify the dynamic region of a prolate cloud with specified initial conditions, we define  a dimensionless quantity - the EUV radiation penetration parameter, which is the ratio of the physical ionising radiation penetration depth to the semi-minor axis of a prolate cloud, 
\begin{equation}
d_{\mathrm{EUV}} = \frac{ \left(\frac{F_{\mathrm{EUV}}}{\alpha_\mathrm{B} \, n^2}\right)}{\left(\frac{a}{\gamma}\right)} = 1.6 \times 10^3 \, \frac{F_{\mathrm{EUV}}^*\, \gamma}{n^2 \, a^*}  (pc)
\label{pene-depth}
\end{equation}
where the major axis $a^{*}$ is in the unit of pc, and $F_{\mathrm{EUV}}^*$ is the EUV ionising radiation flux in units of $10^9$ cm$^{-2}$s$^{-1}$, $\alpha_\mathrm{B}$ is the recombination coefficient of hydrogen ion - election under the 'on-the-spot' approximation \citep{DysonWilliams1997-1} and has the value of $2.0 \times 10^{-13}$ cm$^3$ s$^{-1}$ at a temperature of about $10^4$ K \citep{DysonWilliams1997-1}. This is then taken as a constant, as the equilibrium temperature for ionised material is $\approx 10^4$ K and the dependance of $\alpha_\mathrm{B}$ on temperature is not strong in the region around that temperature.

This dimensionless parameter is comparable in purpose to the dimensionless parameters $\Delta$ and $\Gamma$ (as measures of the overpressure of the ionised gas and ratio of photons re-ionisations/new ionisations respectively), used by \citet{LeflochLazareff1994-1,LeflochLazareff1995-1} for characterisation of an ionisation shock propagation scenario.

In a normal \HII\ region, if the EUV radiation penetration depth is about one hundredth of the minor axis $\frac{a}{\gamma}$, i.e., $d_{\mathrm{EUV}} << 1$, the cloud is in the shock dominated region and would collapse toward the geometrical focus or foci at the final stage of its evolution and we define this mode of the RDI triggered collapse as 'foci convergence'. In this case, an initially spherical cloud would collapse toward the central point of its final structure, and an initially prolate cloud would collapse toward the two foci, the gravitational centres of the cloud. As the value of $d_{\mathrm{EUV}}$ increases, but still much less than 1, the gravitational foci convergence of the cloud is weakened by photo-evaporation,  the cloud collapses toward its major axis. We define this mode of the RDI triggered collapse as 'linear convergence'. Under the very extreme condition of $d_{\mathrm{EUV}} \rightarrow 1 $, the cloud is in photo-evaporation dominant region, and shall totally disperse into its surroundings during its evolution process.  We are only interested in investigating the evolution of the prolate clouds which are not in photo-evaporation dominant region. From our simulation results, we find that $d_{\mathrm{EUV}} $ is a useful diagnostic parameter to indicate the evolution of a prolate cloud under the effect of EUV radiation.

As the distance scales of the ionisation front are generally smaller than an SPH particle smoothing length, it must be treated such that the ionisation front progress through the mass of an SPH particle is tracked, rather than resolved spatially. This is done through the implementation of the grid based method described in Section 3.2.2 of \citet{KesselBurkert2000-1}. The extinction of radiation to a given SPH particle is performed using ray tracing to produce a series of line segment along which the radiation is attenuated. The target SPH particle is then assumed to be a uniform sphere with the radius being defined by its mass and density. The time evolution of the ionisation fraction of each particle is computed from solving the ionisation and recombination equilibrium equation. This allows an ionisation fraction expressed as the equilibrium position of the front \emph{within} the smoothing lengths of the particles.

\section{Results and discussion}
\label{resultsetc}

\citet{NelsonLanger1997-1} investigated the dynamic evolution of three prolate clouds of masses 100, 150 and 200 \msun, subjected to an isotropic interstellar background (FUV) radiation of one Habing unit \citep{Habing1968-1}. All three clouds collapse to a high density spindle at the late stage of the evolution. It is our first interest to investigate what effect an additional plane parallel ionising EUV radiation field would cause on the evolution of these prolate clouds. This is followed by a systematic exploration on the roles played by initial physical and geometrical conditions of a prolate cloud when subject to the same radiation environment.

All subsequent density cross section renders of the simulation data in this paper were produced using the `SPLASH' graphical visualisation tool \citep{splash}.

\subsection{Evolution of high mass prolate clouds}
\label{NelsonLangerSection}
The three clouds under investigation are of same initial density, 100 \htwoden, and axial ratio of $\gamma = 2$. Their properties are listed in Table \ref{highmass}, from which it can be seen that each semi-major axis, $a$, is greater than $a_{\mathrm{crit}}$ indicating that they are supported against purely gravitational collapse.  

We first discuss the evolution of Cloud C  and then describe the general evolutionary features of clouds A, B and C.  The number of SPH particles used in the simulations are 100, 150 and 200K for clouds A, B and C respectively, to satisfy the minimum mass resolution requirement, $10^{-3}$ \msun\ per SPH particle.   
\begin{table}
\centering
\begin{tabular}{lllll}
\cline{1-5}
 Name & Mass (\msun) & $a$ (pc) & $a_{\mathrm{crit}} $(pc) & $\frac{d_{\mathrm{EUV}}}{10^{2}}$ \\
 A  & 100 & 2.68 & 1.58 & 12\\
 B  & 150 & 3.07 & 2.37 & 10 \\
 C  & 200  & 3.38 & 3.16 & 9.5\\
\cline{1-5}
\end{tabular}
\caption{The parameters of the three prolate clouds of similar initial uniform density of 100 \htwoden\ and axial ratio $\gamma =2$. The columns, from left to right, are the name, mass, major axis length $a$, the critical major axis length and the ionising radiation penetration depth parameter (calculated with Equations \ref{majoraxis} and \ref{pene-depth}).}
\label{highmass}
\end{table}
 
\subsubsection{Cloud C - Evolutionary features}
\begin{figure*}
\begin{minipage}{1.0\textwidth}
\center
\includegraphics[width=0.8\textwidth]{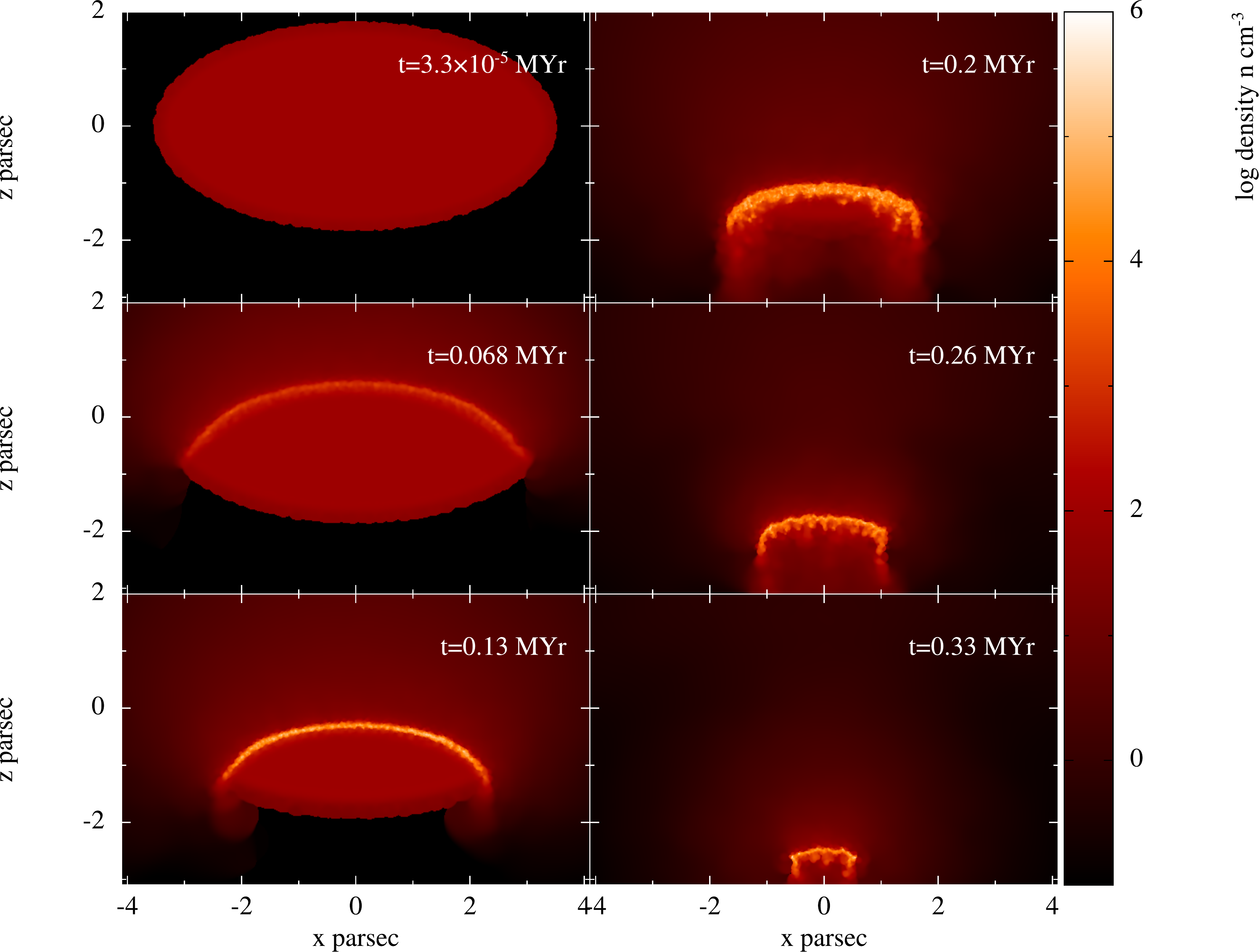}
\end{minipage}
\caption{Sequence of the evolution of the mid-plane cross-sectional number density for Cloud C, of initial density 100 \htwoden\ and $\gamma = 2.0$; subject to an isotropic interstellar background radiation and ionising radiation with the configuration shown in Figure \ref{illustration}. Time sequence is top to bottom then left to right.}
\label{den-mass200}
\end{figure*}

Figure \ref{den-mass200} shows 6 snapshots of the cross sectional number density evolution in the mid-plane for the molecular cloud C over 0.33 Myr. As time progresses from the start of the simulation, the ionisation heating induced shock propagates into the cloud through the upper half ellipsoidal surface (the star-facing side), which is much stronger than that surrounding the lower half ellipsoidal surface caused by FUV only. The shocked thin layer is very distinctive when $t = 0.13$ Myr. At the same time, EUV radiation has photo-evaporated much of the gas material from the  surface of the cloud, such that the overall dimension of the cloud greatly decreases. With the shock propagating into the neural cloud, the condensed thin shell starts to fragment at $t = 0.2$ Myr due to its gravitational instability.  The densities of the gas between the fragments are lower than that in the fragments and are therefore pushed into the cloud by the high pressure in the \HII\ region, to form spike-like microstructures.  These microstructures have higher density than that the neutral interior of the cloud, but would not play significant role over the evolution of the whole system because of their very small volume.  At 0.33 Myr, the remaining material has evolved to a clumpy linear structure of $\approx$ 1pc in length, along which multiple condensed cores are embedded. The peak density increases from $10^{2}$ \htwoden\ at the beginning of the simulation to $\approx 10^{13}$ \htwoden\ at $t = 0.33$ Myr. 

In order to obtain a better impression of the distribution of high density material in the final linear structure, we plot the axial mean density distribution along $x$ axis, similar to the method used by \citet{NelsonPapaloizou1993-1} and \citet{NelsonLanger1999-1}. We divide the length of the prolate cloud along the major-axis into $K$ bins of equal length $2 a /{K} $. We then calculate the mean hydrogen number density for the SPH particles in the bin, i.e., $\bar{n_{i}} = \frac{\sum_{j} n_j }{N_i}$ for the $i^{\mathrm{th}}$ bin along the $x$ axis, with $ i = 1, 2, 3.....K$, $N_i$ is the number of SPH particles of the $i^{\mathrm{th}}$ bin and $j$ is the index of all particles within each bin. $\bar{n_i}$ provides a clear view of the distribution of high density material. Results for the shape of the distribution are converged for a wide range of bin widths relative to smoothing lengths ($20 < K < 2400$ provides identical overall shapes with varying detail for cloud C). $K = 150$ was used for the distributions presented here. 

An alternative implementation using a grid of SPH-style smoothing kernel evaluations with normalised interpolant (such as used by \citet{splash} for the SPLASH plotting tool) plus a density weighting for the $x$-bin average is also investigated for the same purpose as the above. We find the result converging to that by using the simple binning method, albeit at greater time and computational expense. Furthermore, a direct use of the normal SPH kernel average for a grid plus arithmatic average over the grids in an $x$-bin is found insufficient, because it does not highlight the high density regions at all.  
 
We would like to emphasise, however, that it is a qualitative illustration of general high density material distribution, rather than a quantitative representation of core locations and properties, for which the analysis with the core-finding program is used.

For a comparison, we also plot the evolved final axial mean density distribution for the same cloud but without EUV radiation \citep{NelsonLanger1997-1}.  The green lines in the two panels of Figure \ref{NelsonMassCompare} describe the distribution of $\bar{n_{i}}$ along the major-axis ($x$) of cloud C without (upper panel at $t = 3.03$ Myr) and with (lower panel at 0.33 Myr) EUV radiation. An obvious difference which can be seen from these two profiles is that the EUV radiation induced shock could trigger distinctive density peaks along the final linear structure, while most of the less dense material between the cores is blown away by the strong EUV induced photoevaporation. It is worth noting that an apparent high density peak in these plots does not necessarily correspond to a single high density core; multiple cores may be present at different $y$ and $z$ positions within the same $x$-axis bin. In comparison, the FUV only radiation induced shock is much weaker than that of EUV radiation. As such it is about 20 times slower at compressing the gas. Also there are no well separated high density peaks appearing in the FUV radiation only case. We believe that this may be because the FUV radiation is isotropic and the induced weak shock effect is symmetrical about the major axis.

\subsubsection{Clouds A, B \& C - Common and different evolutionary features}
\begin{figure*}
\begin{minipage}{1.0\textwidth}
\center
\includegraphics[width=0.8\textwidth]{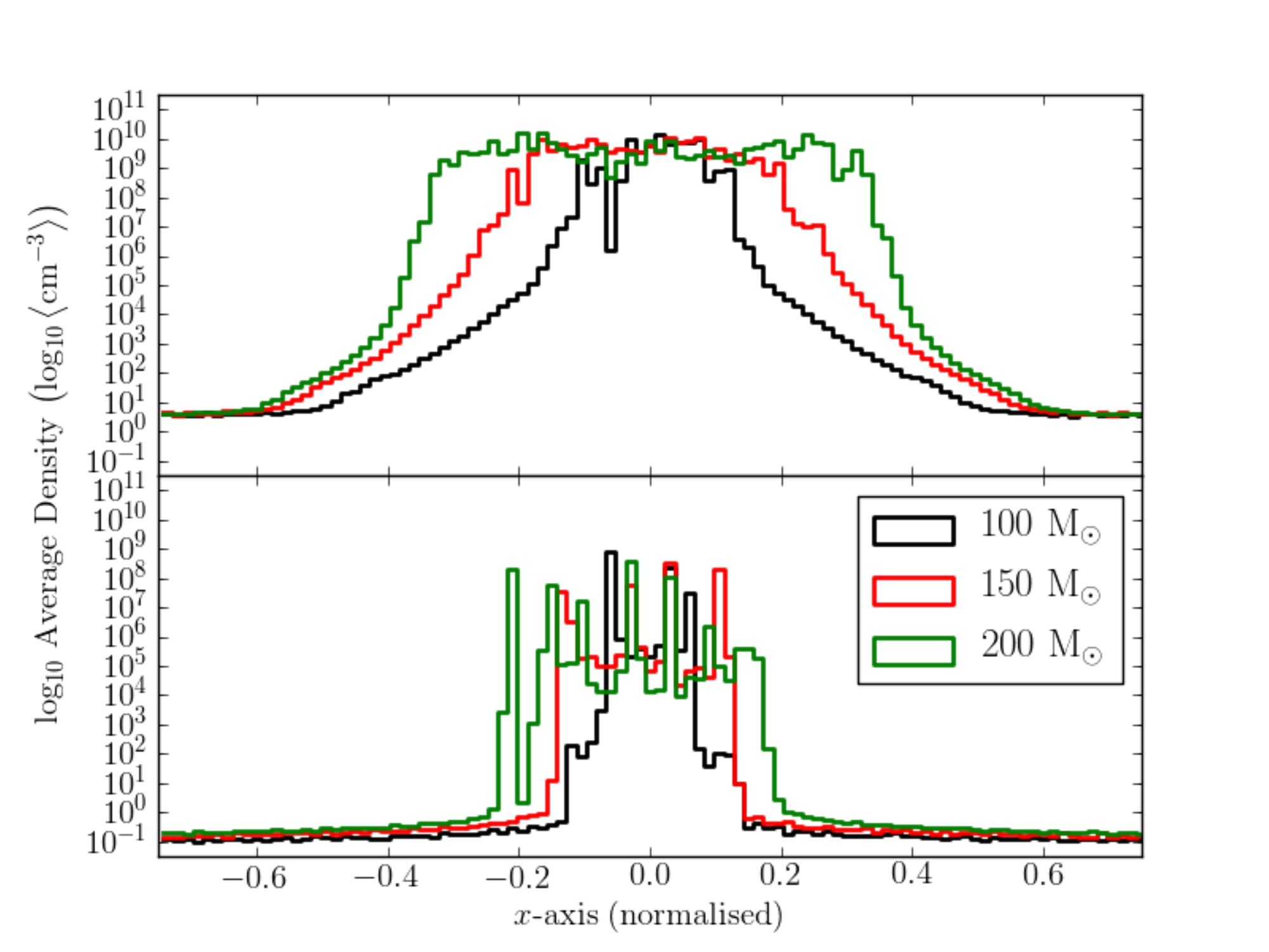}
\end{minipage}
\caption{Mean hydrogen number density along the semi-major axis of three prolate clouds of initial density 100 \htwoden\ and varying masses. Frame (a) is for the three clouds subject to an isotropic FUV radiation only; frame (b) is for the same three clouds subject to both FUV and EUV radiation fields. The time scales for the clouds in frame (a) are 3.10, 3.02 and 3.03 Myr for the clouds of 100, 150 and 200  \msun\ respectively; and 0.33 Myr for all of the clouds in frame (b).}
\label{NelsonMassCompare}
\end{figure*}

For the other two clouds A and B, similar morphological evolution to that of clouds C is observed. Plotted in Figure \ref{NelsonMassCompare} are their axial mean density distributions along the major-axis at the final time step of each simulation. 

It is apparent that the high density cores in all three simulated clouds including EUV radiation scatter over the final clumpy linear structure unlike their corresponding non-EUV simulations. In the FUV-only cases the high density material is more evenly distributed along the final spindle structure. Also apparent in Figure \ref{NelsonMassCompare} is that more gas material remains in the linear structure in the simulation without EUV radiation than that with EUV radiation. This is because the EUV radiation flux is more than 20 times more energetic than the interstellar background FUV radiation, the consequent photoevaporation effect is stronger in a similar proportion. Furthermore, the number of distinctive peaks increases with the initial mass of the cloud. This is understandable as the major axis is longer in higher mass clouds to keep the same initial density, accompanied with an increase in mean mass per unit length. Fragmentation of a longer structure produces more individual fragments.

These highly condensed peaks can be considered to be  potential sites for further star formation. The scattered distinctive high density cores  over the remaining linear structure implies that EUV radiation may be able to trigger a chain of stars to form in the examined prolate clouds at an \HII\ boundary in less than 0.5 Myr. In contrast, the same prolate clouds further away from a massive star are more likely to form a condensed filamentary structure under the effect of the FUV only radiation over a period of a few Myr.


\begin{table}
\centering
\begin{tabular}{llllll}
\cline{1-6}
 Name & Density & $a$ & Flux & $d_{\mathrm{EUV}}$ & Time \\
 & {(\htwoden)} & {(pc)} & {(cm$^{-2}$ s$^{-1}$)} & {($10^{-2}$)} & {(Myr)}\\
 C  & 100 & 3.38 & $10^{9}$ & 9.5 & 0.334 \\
 D1 & 400   & 2.13 & $10^{9}$ & 0.93 & 0.327 \\
 D2 & 600   & 1.86 & $10^{9}$ & 0.48 & 0.288 \\
 D3 & 1,200 & 1.48 & $10^{9}$ & 0.15 & 0.233 \\
 E1 & 100   & 3.38 & $10^{7}$ & 0.095 & 0.966 \\
 E2 & 100   & 3.38 & $10^{8}$ & 0.95 & 0.776 \\
 E3 & 100   & 3.38 & $8 \times 10^{9}$ & 76 & 0.268 \\
\cline{1-6}
\end{tabular}
\caption{The parameters of the further exploration high mass prolate clouds. The columns, from left to right, are the name, density, major axis length $a$, incident EUV flux and the ionising radiation penetration depth parameter (calculated with Equations \ref{majoraxis} and \ref{pene-depth}). Time indicates the final simulation time. Cloud C is the same as in the previous section; clouds D1-3 are the additional density-varied tests, and clouds E1-3 are the additional incident flux-varied tests.}
\label{additionaltable}
\end{table}

\subsubsection{D Series - Effects of varied initial density}

In order to inspect how the evolutionary destiny would change if the initial density of the above prolate cloud is increased, we investigated the evolution of another three prolate clouds which has same mass of 200 \msun, but different initial densities. These tests are labeled as D1-3 in Table \ref{additionaltable}. \label{Dtests}

\begin{figure}
\center
\includegraphics[width=0.53\textwidth]{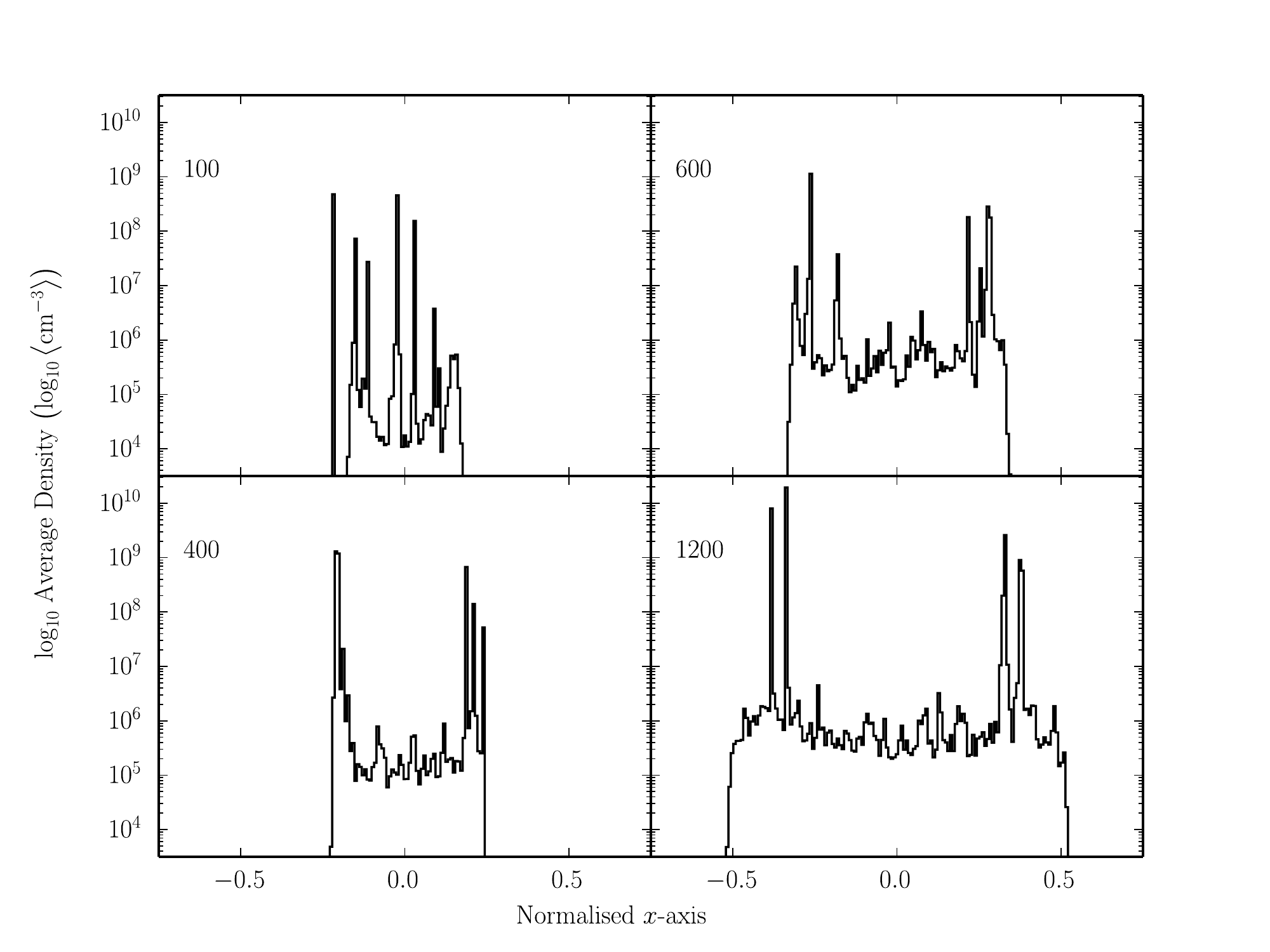}
\caption{The axial mean number density profile at the final timesteps in four molecular clouds of same mass of 200 \msun, and $\gamma = 2$, but different initial densities as shown in the top-left corner in each panel. The densities are in unit of \htwoden. The times at which the simulations ended are shown in Table \ref{additionaltable}.}
\label{axialden-highM-diffden}
\end{figure}

Presented in Figure \ref{axialden-highM-diffden} are the axial mean density distributions along the formed linear structures at the final timestep. The plot for cloud C is also plotted for comparison. It is interesting to see that with increasing initial density, the condensed cores gradually move toward the two foci or say the two ends of the final filament structure. This is because the ionising radiation penetration depth parameter decreases with the increase of the initial density, the mode of the evolution of the cloud changes from linear to foci convergence.  

\subsubsection{E Series - Effects of varied EUV Flux}

\begin{figure}
\center
\includegraphics[width=0.53\textwidth]{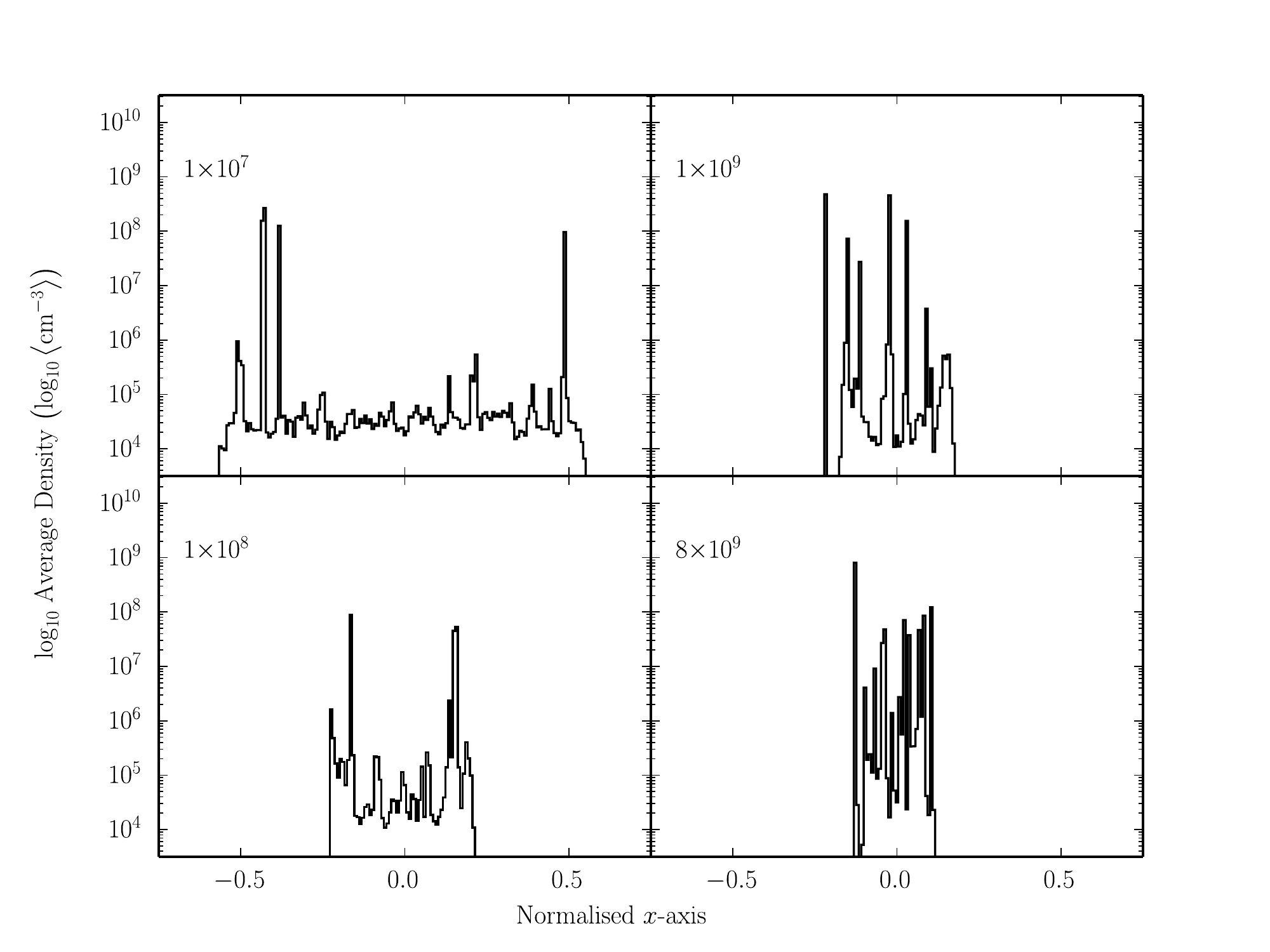}
\caption{The axial mean number density profile at the final timesteps in four molecular clouds of same mass of 200 \msun, $\gamma = 2$ and initial density of 100 \htwoden,  but under the effect of different EUV radiation fluxes as shown in the top-left corner in each panel. The fluxes are in units of cm$^{-2}$s$^{-1}$. The times at which the simulations ended are shown in Table \ref{additionaltable}.}
\label{axialden-highM-diffEUV}
\end{figure}

The four panels in Figure \ref{axialden-highM-diffEUV} illustrate the axial mean density distributions, in simulations with a cloud of the same initial conditions as cloud C, but different EUV radiation fluxes; notated as clouds E1-3 in Table \ref{additionaltable}. It is seen that the distribution of the condensed cores gradually changes from the two-foci concentrated to scattered over the whole filament, with the increase of the EUV radiation flux, i.e., increase of d$_{\mathrm{EUV}}$.  The mode of the evolution of the cloud changes from foci to linear convergency. \label{Etests}

Next, we turn to a systematic investigation on the evolutionary features of prolate clouds of an intermediate initial mass of 30 \msun \,, a typical initial density around $10^3$ \htwoden\ and different initial shapes.

\subsection{Evolution of prolate clouds of 30 \msun}
\label{perpendicularsetup}
 The prolate clouds in this investigation have masses of 30 \msun, different initial densities of 600 and 1200 \htwoden, and varied initial geometrical shapes defined by the axial ratio parameter $ 1 \le \gamma \le 8$.  We categorise  them into two groups G1 and G2, as listed in Table \ref{perpend_tests_table}.  Their initial major axes are all larger than their $a_{\mathrm{crit}}$, which means they are all stable against purely gravitational collapse. There are 19 clouds in each group and are numbered from 1 to 19. The identification for each cloud is notated as G1(No.) and G2(No.), e.g. the 5th cloud in the G2 series is named as G2(5). Each of the simulations for the 38 clouds was run with $10^5$ SPH particles, leading to a mass resolution of $3.0 \times 10^{-4}$ \msun\ per SPH particle, a higher resolution than required by the convergence tests ($10^{-3}$ \msun\ per SPH particle).

In the following, we present the simulation data and analyse the features of the evolutionary sequence for the two groups of prolate clouds.
\begin{table}
\centering
\begin{tabular}{lllllll}
\cline{1-7} 
 & & & G1 & & G2 & \\
No. & $\gamma$ & $a_{\mathrm{crit}}$ & $a_{600}$ & $\frac{d_{\mathrm{EUV}}}{10^{2}}$ &$a_{1200}$  & $\frac{d_{\mathrm{EUV}}}{10^{2}}$  \\
\cline{1-7} 
1 & 1.000 & 0.052 & 0.623 &0.713 & 0.494  & 0.225  \\
2 & 1.250 & 0.060 & 0.722 &0.770 & 0.573 &0.242  \\ 
3 & 1.500 &  0.067 & 0.816 & 0.817 & 0.648  &0.257 \\ 
4& 1.750 &  0.073& 0.904 & 0.860 & 0.718  &0.271 \\ 
5 & 2.000 &  0.079& 0.988 &0.900 &0.784  & 0.283\\ 
6 & 2.250 &  0.084& 1.069 & 0.935 & 0.849  & 0.294 \\ 
7 & 2.500 &   0.039& 1.147 & 0.969& 0.910  & 0.305\\
8 & 2.750 &  0.093& 1.222  & 1.001&  0.970  & 0.315\\ 
9 & 3.000 &  0.097& 1.295 & 1.030 & 1.028   & 0.324 \\
10& 3.250 & 0.101& 1.366  & 1.057&1.084    & 0.333\\
11 & 3.500& 0.104& 1.435  & 1.084   &1.139   &0.341\\
12 & 3.750 &0.107& 1.503  & 1.109 &1.193  & 0.349\\
13 & 4.00 & 0.110& 1.569  & 1.133& 1.245 & 0.357\\
14  & 4.50 &0.116& 1.697 & 1.179 &1.347  & 0.371\\
15 & 5.00 & 0.121& 1.820  & 1.221&1.445 & 0.384 \\
16 & 5.50 &0.126& 1.940  & 1.260 &1.540  &0.397 \\
17 & 6.00 &0.130& 2.056  & 1.297 &1.632  &0.408 \\
18 & 7.00 & 0.138& 2.278  &1.366  &1.808 & 0.430\\
19 & 8.00 & 0.145& 2.490  & 1.428 & 1.977 & 0.450\\
\cline{1-7} 
\end{tabular}
\caption{Parameters of two groups of molecular clouds of same mass but different initial densities of 600 and 1200 \htwoden.  From left to right, columns 1-3 are the number identity,  axial ratio and the critical semi major axis  defined by Equation \ref{majoraxis} for both  G1 and G2 clouds. Columns 4-5 are the major axis and $d_{\mathrm{EUV}}$ defined by Equation \ref{pene-depth} for the G1 clouds, and columns 5-6 are the same parameters for the G2 clouds.  All of the semi-major axes and critical semi-major axes are in units of pc and the penetration depth is unitless.}
\label{perpend_tests_table}
\end{table}

\subsubsection{G2 series - Effects of varied initial geometry}
\label{G2}
The morphological evolution of the clouds in the G2 group are very similar to each other, so we only describe in detail the evolutionary sequence for the cloud G2(5), which is of an initial axial ratio of 2 and an initial density 1,200 \htwoden. Then, we have a general description of the evolutionary features of the whole group.

\begin{figure*}
\begin{minipage}{1.0\textwidth}
\center
\includegraphics[width=0.8\textwidth]{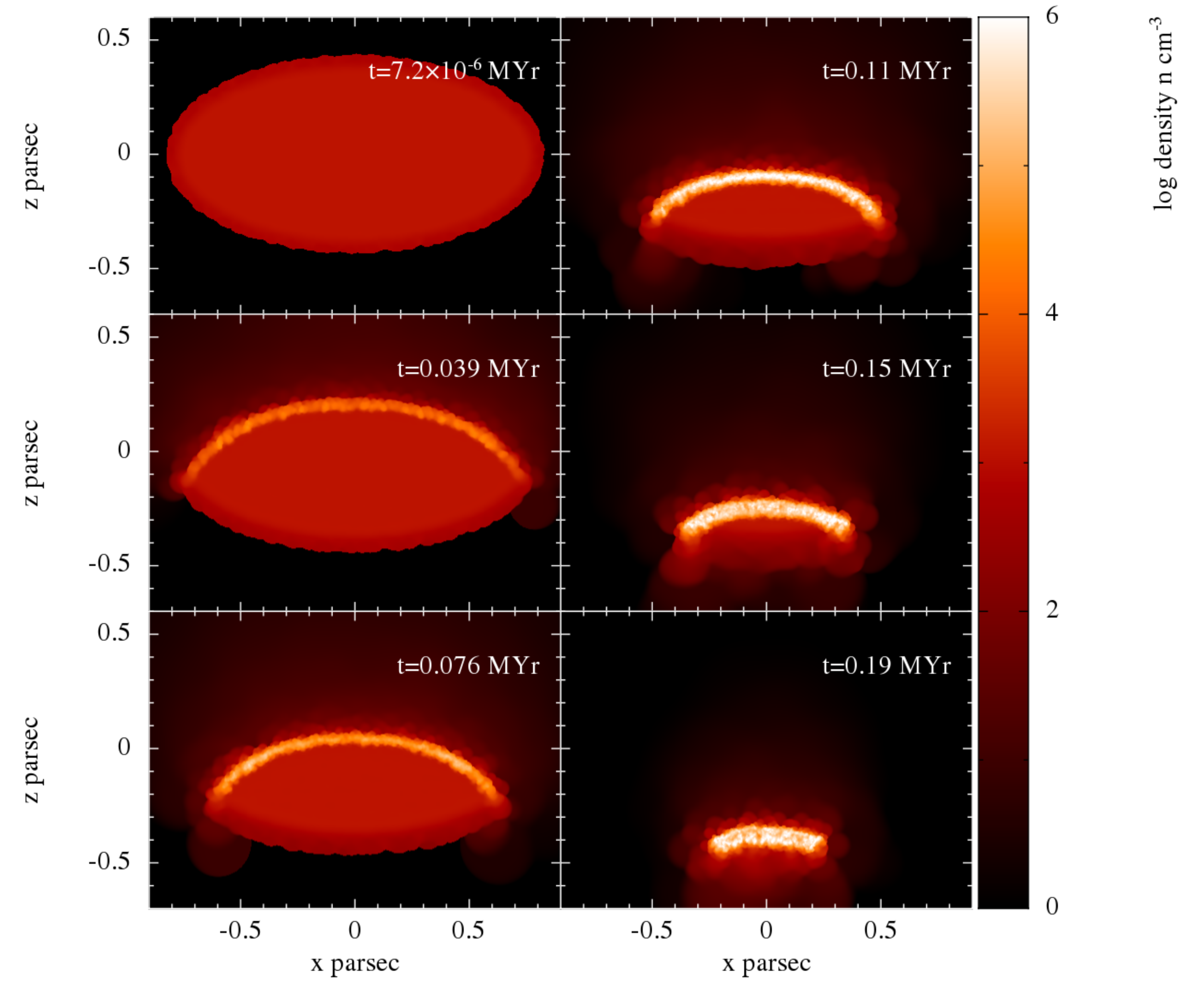}
\end{minipage}
\caption{Evolution of the cross-sectional density in the mid-plane for the prolate cloud G2(5) over 0.19 Myr. Time is displayed in the upper right of each panel. The order of time evolution is top to bottom and then left to right.}
\label{evo-1200}
\end{figure*}

\begin{figure*}
\begin{minipage}{1.0\textwidth}
\center
\includegraphics[width=0.8\textwidth]{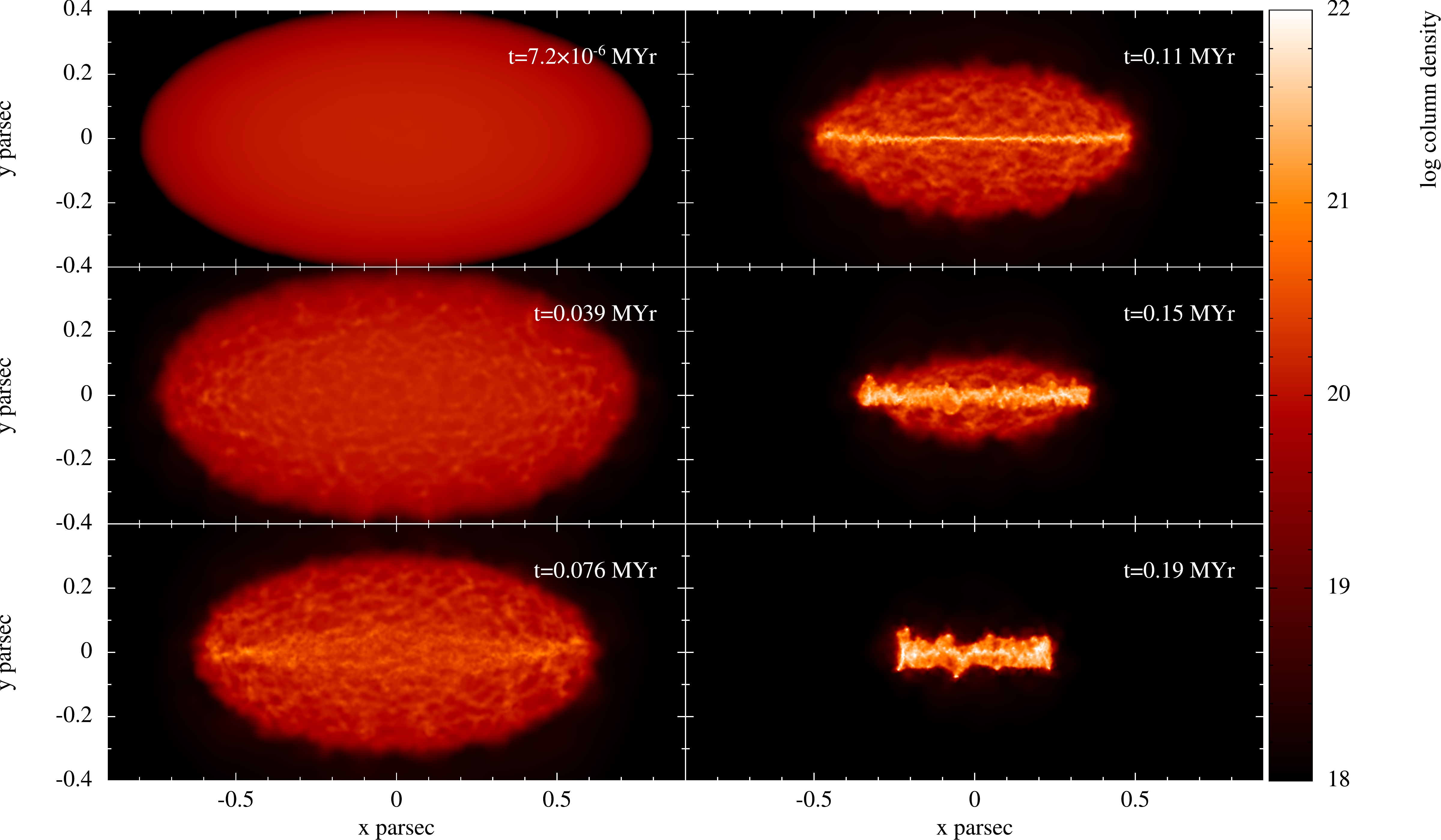}
\end{minipage}
\caption{Evolution in the $xy$-plane projection (overhead column density) of the prolate cloud G2(5). Time is displayed in the upper right of each panel. The development of a linear filament-like structure along the `spine' of the cloud can be seen to have occured by 0.11 Myr; its fragmentation into zig-zagging, thinner filaments visible at the final time, 0.19 Myr.}
\label{overhead-1200}
\end{figure*}

The six panels in Figure \ref{evo-1200} describe the evolution of the cross-sectional number density in the  mid-plane ($x-z $ and $y = 0$ ) of the cloud G2(5). The morphological evolution appears to follow the general picture described by the RDI mechanism. A condensed gas layer at the upper ellipsoidal surface has formed within 0.11 Myr. The density inside the shocked layer increases as it propagates inwards. At $t \approx 0.19$ Myr, the highly condensed layer fragments, creating a curved clumpy filamentary structure with condensed cores embedded. 
The corresponding overhead ($x-y$) view of the evolution of the cloud displayed in Figure \ref{overhead-1200}, further confirms the formation of the filamentary structure and its fragmentation. It is seen that    
the filamentary structure forms as a high-density `spine' aligned with the semi-major axis at $t = 0.11$ Myr. The material in the two hemispheres is seen converging to the major x-axis, and material from negative $y$ heads toward the positive $y$ direction, and vice-versa. At $t = 0.15 $ Myr, this thin and long structure starts fragmentation. Some of the fragments disperse off the major axis, and a broadly zig-zag fragment-core structure is left at $t = 0.19$ Myr. 

\begin{figure}
\center
\includegraphics[width=0.47\textwidth]{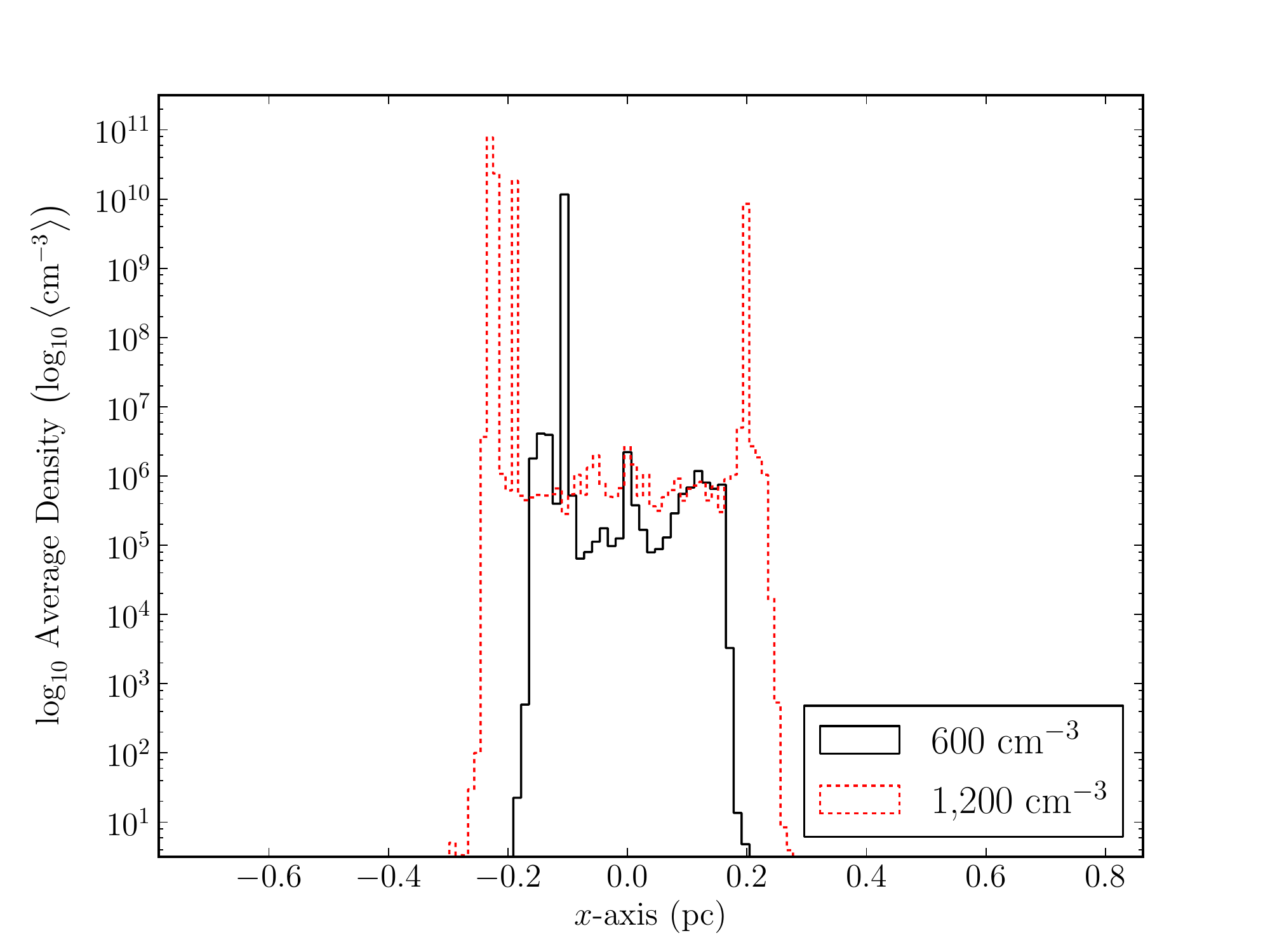}
\caption{The axial mean number density profile for clouds G1(5) (600 \htwoden, $\gamma=2$) in solid black and G2(5) (1,200 \htwoden, $\gamma=2$) in dashed red. G1(5) is at $t = 0.22$ Myr and G2(5) at $t = 0.19$ Myr.}
\label{meanden6001200}
\end{figure}

The red line in Figure \ref{meanden6001200} describes the axial mean density distribution at $t = 0.19$ Myr. It shows distinctive high density peaks forming at the two ends of the filamentary structure. The penetration depth parameter of cloud G2(5) is $d_{\mathrm{EUV}} = 0.283\%$ as shown in Table \ref{perpend_tests_table}, which means that EUV radiation induced shock dominates the evolution of G2(5) and enhances the self-gravity of the cloud G2(5) so that most of remaining condensed gas is driven toward the two foci of the cloud over its evolution (foci convergency). 
 
\begin{figure*}
\begin{minipage}{1.0\textwidth}
\center
\includegraphics[width=0.8\textwidth]{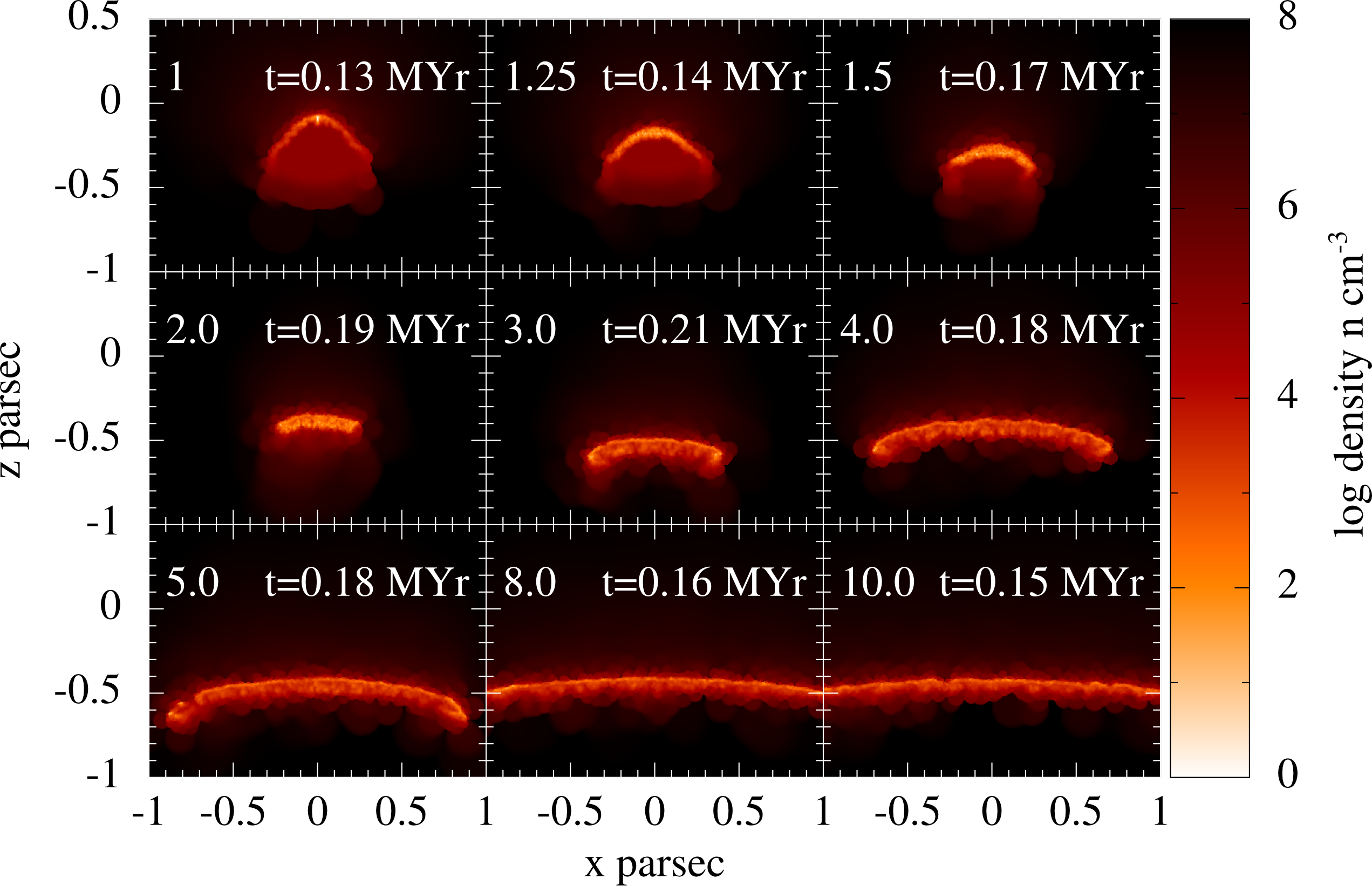}
\end{minipage}
\caption{The cross-sectional density in the mid-plane for the 9 representative prolate clouds in the G2 series at their final time step. The number in the upper-left of each panel is the axial ratio $\gamma$, and the number in the upper right is the evolutionary time.}
\label{final-1200}
\end{figure*}

\begin{figure*}
\begin{minipage}{1.0\textwidth}
\center
\includegraphics[width=1.0\textwidth]{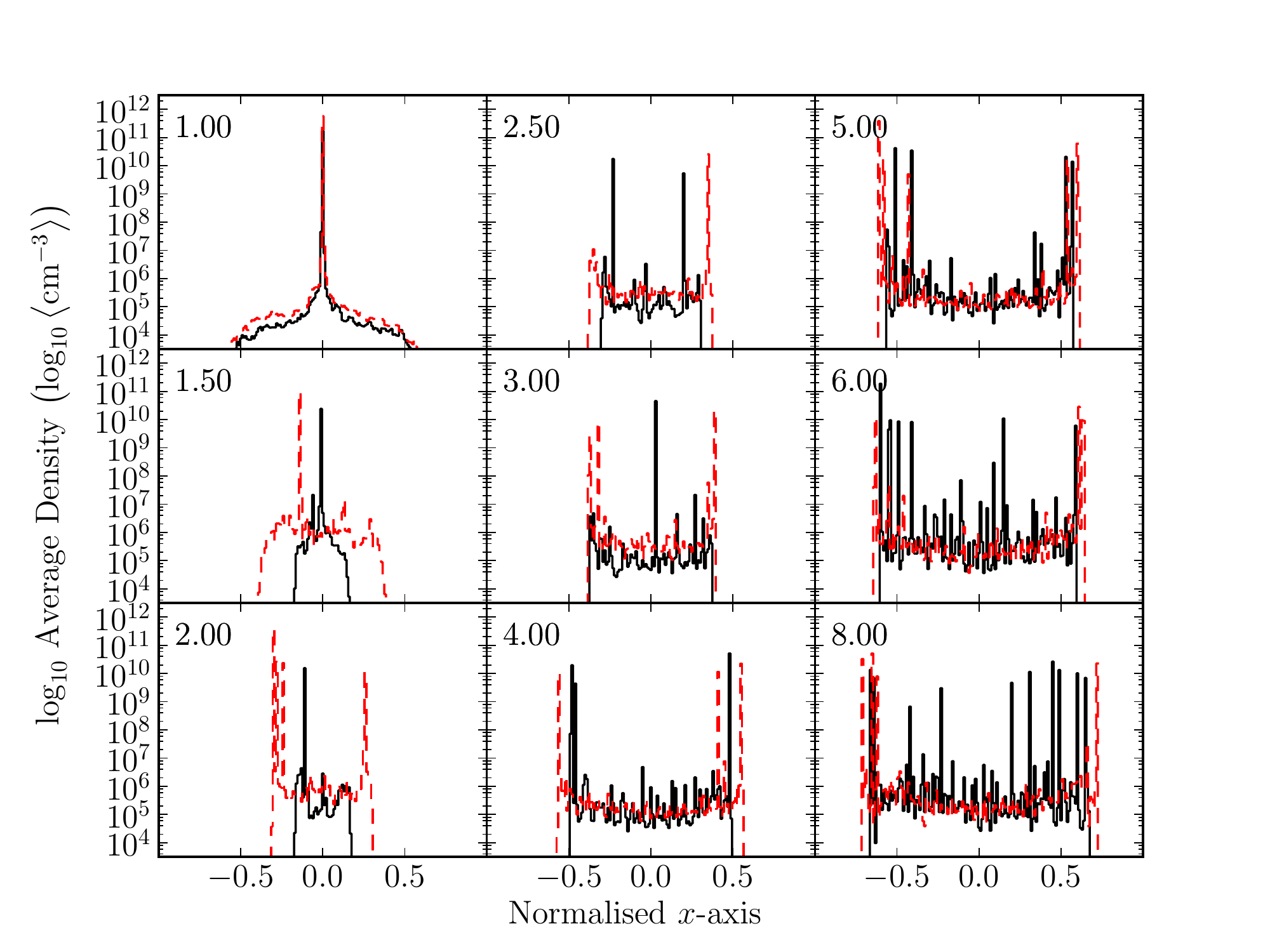}
\end{minipage}
\caption{The axial mean density over $x$-axis (normalised to the initial cloud semi-major axis) for nine of the G1 (black solid line) and G2 (red dashed line) series clouds. The number in the upper left of each panel is the $\gamma$ value of the clouds.}
\label{axial-den-all}
\end{figure*}

Figure \ref{final-1200} describes the cross sectional density distribution in the mid-plane for nine representative clouds in the G2 group, at the end of the simulation.  We can see that the curvature of the final morphology of the formed structure decreases with the the increase of the initial axial ratio $\gamma$. An initially spherical cloud ($\gamma = 1$) evolves to a type B BRC with a highly condensed core forming at its head in the shortest time of 0.13 Myr.  Clouds of $ 1 < \gamma \le 1.5$ evolve to a type A BRC with a dense core forming at its head in longer time between 0.13 and 0.17 Myr. In the clouds of $1.5 < \gamma \le 3$, a bright rimmed curving and clumpy filament forms with cores being embedded along the linear structure in an increasing timescale up to 0.21 Myr. It is seen that the core collapsing time of a cloud increases when the gravitational centre changes from one to two.  However, in the clouds of initial $  3 < \gamma \le 10.0 $, the core collapsing time decreases from 0.21 to 0.15 Myr with  $\gamma$. This may be because the initial cloud having $\gamma > 3$ becomes more and more elongated,  with $\gamma$ increasing, the converging gas material has shorter and shorter distance (therefore shorter converging time) to travel to collapse toward their two foci. Therefore spherical and highly ellipsoidal clouds have shorter core formation  than those of the mid range of axial ratios (this can also be seen in Figure \ref{time_vs_ratio}).

It is also of interest to look at the axial mean density profiles of a set of clouds in the G2 group.  The red dashed lines in Figure \ref{axial-den-all} reveal the location of the high density peak(s) triggered by the EUV radiation flux for 9 clouds selected from the G2 group (the axial ratio for each is displayed in the upper left of the panel). An initially spherical cloud G2(1) converges to its gravitational centre to form a single dense peak as shown in the first panel. As axial ratio increases, high density cores are forming at the two ends of the final structures. The G2 group clouds all have $d_{\mathrm{EUV}} < 0.5\%$ and therefore all collapse in the mode of foci convergency.         

\subsubsection{G1 series - Effects of varied initial geometry with halved initial density}
\label{G1}
The clouds in the G1 series have an initial density of half that of the G2 group, 600 \htwoden. Their EUV radiation flux penetration parameters are in the range $0.713 \le d_{\mathrm{EUV}} < 1.428 $ \%, larger than that of all of the clouds in G2 group.   

The morphological evolution of G1 group clouds are similar to that of G2 group clouds. Clouds of $\gamma$ equal or close to 1 form type B or type A BRCs with a single core forming at the its head.  As $\gamma$ increases, the clouds evolve into filamentary structures with cores embedded inside. For even higher axial ratios, warm but dense capillary structure appears ahead of the shocked layer as seen in Cloud C as well. The axial mean density profiles for 9 G1 group clouds are plotted as black solid lines in Figure \ref{axial-den-all}, which describe a coverage of all dynamic features of the G1 series.

As seen from Figure \ref{axial-den-all}, the clouds of $ 1 \le \gamma < 2.0 $ and $ 0.713 \le   d_{\mathrm{EUV}} < 0.9\%$ in both groups are spherical or quasi-spherical and  evolve to similar structures with a highly condensed core, except more gas material is evaporated from G1 group clouds compared to the G2 group clouds. This is shown by the narrower density profile when compared with the G2(1-4) clouds. The above feature can be explained by the higher values of $d_{\mathrm{EUV}}$ in G1 clouds, where more surface material is photoevaporated. However, the overall dynamical evolution of these clouds can still be categorised as shock dominant, as most of remaining material in the cloud converges to the gravitational centre of the BRCs.

Clouds of $ 2 \le \gamma < 6.0 $ and $ 0.90\% \le   d_{\mathrm{EUV}} < 1.26\%$ in the G1 group not only develop highly condensed cores at one or both ends of the final filamentary structure, but also between the two foci, especially the middle core in the cloud G1(9) of $\gamma =3 $ has a much higher mean density then the sides cores in the same cloud. The above feature suggests that the EUV radiation induced shock dominance decreases. As such, the gravitational convergence toward the two foci is gradually weakened and more gas collapses toward the major axis to form a filament, which then fragments into a few dense cores. It appears that their collapse modes are in a transition region between foci convergency and linear convergency.
 
The clouds having axial ratios $6.0 \le \gamma \le 8.00$ and $d_{\mathrm{EUV}} > 1.26\%$ all collapse in the mode of linear convergency, and the condensed cores spread over the final filamentary structure. For example, in the cloud of $ \gamma = 8.00 $ and $d_{\mathrm{EUV}} = 1.43$\%, convergence toward two foci has broken, the high density cores have similar mean peak density as the consequence of the fragmentation of the final filamentary structure. 

\subsubsection{Effects of varied cloud ratio and lower densities}
\label{G0}
To confirm the correlation observed between $d_{\mathrm{EUV}}$ and the evolutionary destiny of a cloud, two additional sets of simulations were run with prolate clouds of 30 \msun, but of lower initial densities, 300 and 100 \htwoden.  Each group has four different clouds of $\gamma = 1.5, 2.0, 2.5$ and $4$. With these initial conditions, the 300 \htwoden\ clouds have an ionising depth parameter of $ 2.6 \% \le d_{\mathrm{EUV}} \le 3.6\% $, and for the clouds of 100 \htwoden, $ 16\% \le d_{\mathrm{EUV}} \le 22\%$. In total 8 simulations were run with the same mass resolution as used in the G1 and G2 series simulations.  

The morphological evolution and the axial mean density profiles are qualitatively similar to that of the highly ellipsoidal clouds in the G1 simulations, so we do not present similar plots to Figures \ref{final-1200} and \ref{axial-den-all}. None of them collapse in the mode of foci convergency. We  select a representative from the 8 simulations to compare its mode of convergence with that of the G2(5) and G1(5) clouds illustrated in Sections \ref{G2} and \ref{G1} respectively. The cloud with initial density of 100 \htwoden, $\gamma =2 $ and $d_{\mathrm{EUV}} = 17.8 $ is chosen, and will be notated as cloud G0.

\begin{figure}
\center
\includegraphics[width=0.47\textwidth]{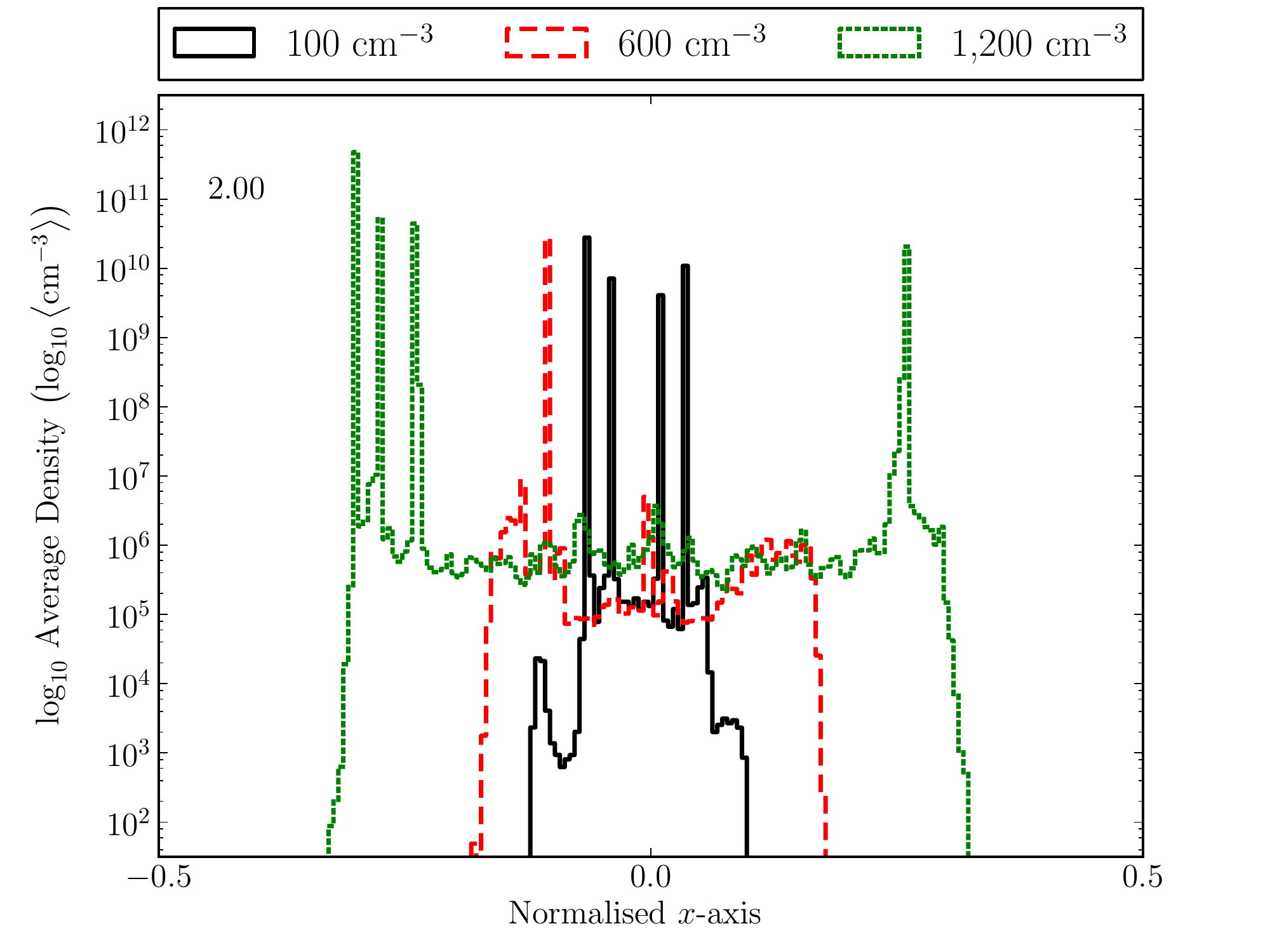}
\caption{The axial mean number density distribution for molecular clouds of 30 \msun\ and $\gamma=2$ but different initial densities. The legend boxes describe the line style and colour correspondence for the densities of the featured clouds, G0 (100 \htwoden), G1(5) (600 \htwoden) and G2(5) (1,200 \htwoden). The evolutionary times of the clouds featured are 0.25, 0.22 and 0.19 Myr for G0, G1(5) and G2(5) respectively.}
\label{meanden1006001200}
\end{figure}

Figure \ref{meanden1006001200}  shows a comparison of the axial mean density profile over the normalised $x$ axis for three molecular clouds of $\gamma =2$, $M =30 $\msun\, and different initial densities of 100 (black line for G0), 600 (red for G1(5)) and 1,200 (green for G2(5)) \htwoden. It is clearly seen again that the mode of collapse in the three clouds changes from linear convergency in G0, to foci-linear mixture convergency in cloud G1(5), then to foci convergency in G2(5), with $d_{\mathrm{EUV}}$ decreasing from 17.8\% to  0.28\%.  

Table \ref{summary} presents a summary on the evolutionary destiny of all the investigated clouds in this series, related to the diagnostic parameter $d_{\mathrm{EUV}}$.

\begin{table}
\centering
\begin{tabular}{ccccc}
\cline{1-5}
 Cloud& Mass  & n  & $d_{\mathrm{EUV}}$ & mode of \\
 Name & (\msun) & (\htwoden) & (\%)   & convergent \\
   G0 & 30 & 100 & 17.8 & linear \\
 G1(17-19) & 30 & 600  &1.3 - 1.43  &  linear \\
 G1(5-16)&30&600 &0.90-1.26 & foci/linear \\
 G1(1-4) &30&600& 0.71 - 0.86 & foci \\ 
 G2(1-19) & 30 & 1200 & 0.23 - 0.45 & foci\\
\cline{1-5}
\end{tabular}
\caption{A summary of the evolutionary destiny of the molecular clouds of mass 30 \msun.}
\label{summary}
\end{table}

\subsubsection{The location of cores}
From the above investigation, it is known that high density cores formed in clouds of lower $d_{\mathrm{EUV}}$ ($\le 1.25$\%) tend to locate around the two ends (foci) of the final filamentary structure. Now investigated are the detailed location profiles of condensed cores along the $x$-axis using the core finding program described in Section \ref{corefinder}. For this objective, we are only interested in the cores with a peak density ($n_c$) higher than $10^8$ \htwoden, which can be taken as the potential seeds for new stars to form \citep{NelsonLanger1997-1}. In each panel the short horizontal lines specify the initial extent of the semi-major axis along the $x$-direction for clouds of the  $\gamma$ and initial density specified specified in the plot. 

\begin{figure*}
\begin{minipage}{1.0\textwidth}
\center
\includegraphics[width=1.0\textwidth]{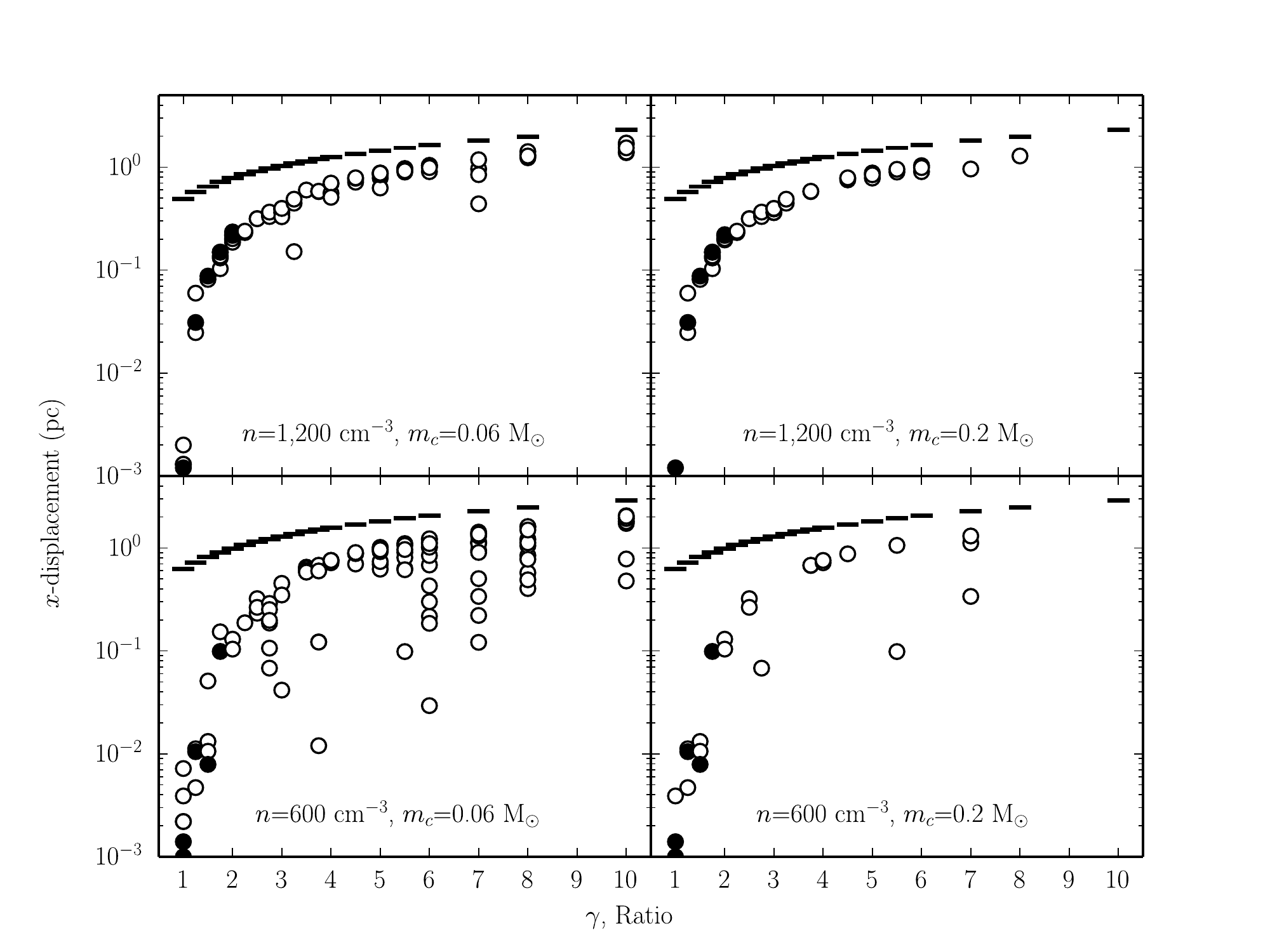}
\end{minipage}
\caption{$x$-displacement locations ($\left| x_c \right|$) of condensed cores formed in the G1 and G2 cloud series with different criteria on the threshold of core mass $m_{\mathrm{c}}$.  $n$ is the initial density of the clouds whose cores are sampled in the corresponding panel. The short horizontal line for each $\gamma$ examined denotes the initial semi-major axis of the cloud. The white circles indicate cores of density $10^{8} \le n_{\mathrm{c}}  \le 10^{12}$ \htwoden\ and the black circles core densities of $ n_{\mathrm{c}} \ge 10^{12} $ \htwoden.}
\label{disp-panels}
\end{figure*}

Plotted in Figure \ref{disp-panels} are the distributions of condensed cores in G1 and G2 clouds of different $\gamma$. The derived data of cores from the corefinding process has been further filtered by the peak density, $n_c$, ($ \ge 10^8$ \htwoden\ in all cases described here) and a minimum mass threshold $m_c$. The two panels on the left have a selection of $m_c \ge 0.06$ \msun, and the two on the right of $m_c \ge 0.2$ \msun. The $x$-displacement parameter is the modulus of the $x$-axis position of the peak of the core, $\left| x_c \right|$. Within each panel, two peak density regimes are distinguished by white filled circles, indicating a density of $ 10^{8} \le n_c \le 10^{12}$ \htwoden, and black filled circles for $n_c > 10^{12}$ \htwoden, being cores of extremely high density. It is seen from each panel in Figure \ref{disp-panels}, that extremely high density cores only form in the clouds of lower $\gamma$ values and appear at the focus points.

In the two panels on the left, where $m_c \ge 0.06$ \msun, the upper of these is for the G2 series of clouds, and the lower for the G1 series. With increasing $\gamma$, high density cores form mainly around the foci of the ellipsoidal cloud in G2, but appear scattered over the whole cloud length in some of the G1 clouds.

The results for the higher core mass criteria ($m_c > 0.2$ \msun) are presented in the two panels on the right of Figure \ref{disp-panels}. The upper of these being the G2 series of clouds, and the lower the G1 series. It is seen that the high mass cores in all clouds of the G2 series are located at the centre or foci along the $x$-axis. The picture is not so simple in G1 clouds. In clouds of $\gamma \le 2$, high mass cores appear close to the centre point $x =0$. Some clouds of $\gamma > 2$  have the high mass cores at two foci and closer to the middle of the $x$-axis as well. Some more ellipsoidal clouds have high mass core(s) either at/around the foci or spread between the foci. A few of the higher $\gamma$ clouds in G1 have no core(s) with mass higher than 0.2 \msun.

The general picture that is revealed is  that the clouds in the G2 and the low $\gamma$ clouds in the G1 groups have almost all of their cores located around their foci.  Clouds of higher $\gamma$ values in G1 have their condensed cores spread along the $x$ axis. The different core distributions between these clouds can be explained by the lower d$_{\mathrm{EUV}}$ in the G2 and low-$\gamma$ G1 clouds compared to the high $\gamma$ G1 clouds.

\subsubsection{The total core mass and core formation time}
In order to evaluate the efficiency of EUV radiation triggered potential star formation in the different prolate clouds of the G1 and G2 groups, we compare the total mass of dense cores and the time for high density core formation in clouds of different $\gamma$ in both groups.
      
\begin{figure}
\center
\includegraphics[width=0.47\textwidth]{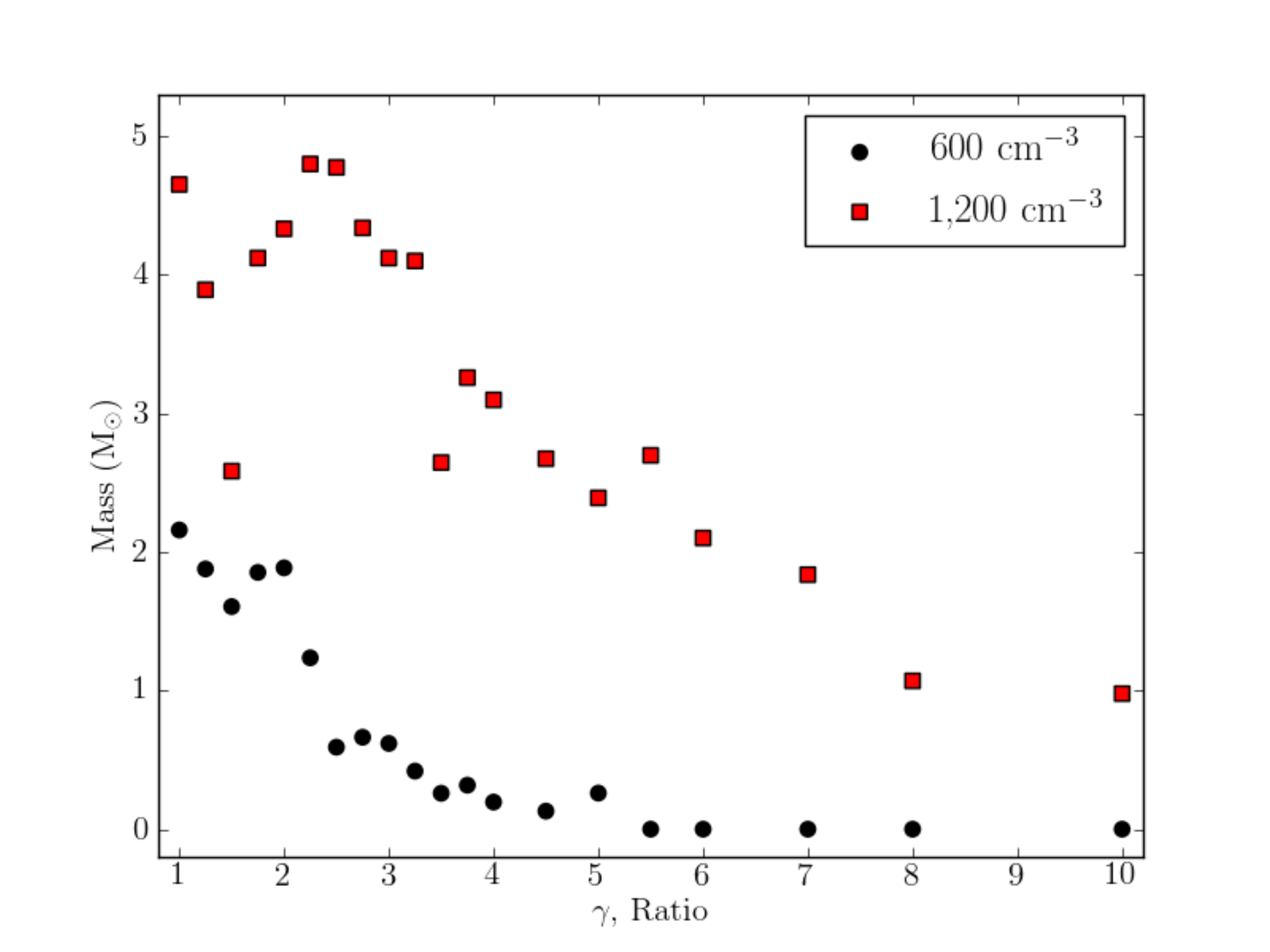}
\caption{The total mass $m_{\mathrm{tot}}$ of all cores with selection criteria of  $n_c \ge 10^{6}$ \htwoden\ and $m_c=0.06$ \msun\ at the end of the simulated evolution for all G1 and G2 clouds.}
\label{totmass_gamma}
\end{figure}

Plotted in  Figures \ref{totmass_gamma} is the variation of the total mass of high density cores in a cloud, $m_{\mathrm{tot}}$, with $\gamma$ for  both groups. It is clearly seen that for each pair of G1 and G2 clouds of the same $\gamma$, $m_{\mathrm{tot}}$ for the G2 cloud is more than double that of the G1 cloud. G1 group clouds have higher $d_{\mathrm{EUV}}$ and therefore lose more material through photoevaporation. The range of $m_{\mathrm{tot}}$ is 1 - 4.85 \msun\ in the G2 series and 0.05 - 2.2 \msun\ in the G1 series. 

However the variation of $m_{\mathrm{tot}}$ over $\gamma$ in each individual group is non-monotonic. Taking G2 group as an example, the spherical cloud has highest degree of convergence, so it has the maximum total core mass. When $1 < \gamma \le 1.5 $, although the cloud becomes an ellipsoid, the two foci are still very close to each other that their effect on gathering gas toward them is similar to one focus cloud. This can be confirmed by the single high density peak in the corresponding axial mean density distribution (in red lines) in Figure \ref{axial-den-all}.  The final structure still keep the morphology of a single BRC as shown in Figure \ref{evo-1200}.  When $\gamma$ increases to 1.5, the overall gravitational convergence toward the centre of mass decreases, so the total core mass $m_{\mathrm{tot}}$ of the high density core decreases with $\gamma$, to the value of 2.6 \msun. 

When $1.75 \le \gamma \le 2.25$, the distance between the two foci in a cloud increases to such a degree that two foci convergency becomes    obvious, as shown in the corresponding panels (in red lines) in  Figure \ref{axial-den-all}, the morphology of the final cloud is no longer a single BRC but a linear structure as shown in Figure\ref{evo-1200}.  Now there are two gravitational converging centres to accrete gas, so the total core mass shows a sharp increase to 4.2 \msun in cloud having $\gamma = 1.75$, then slightly increases with $\gamma$ up to 4.85 \msun in the cloud having $\gamma = 2.25$.   

With further increase  in $\gamma$, the initial cloud becomes more and more elongated, the initial mass per unit length along the major axis become lower and lower and the gas available to be accreted by the two foci gets less and less.  Therefore $m_{\mathrm{tot}}$ decreases with $\gamma$ just as shown in Figure \ref{totmass_gamma}.   

\begin{figure}
\center
\includegraphics[width=0.47\textwidth]{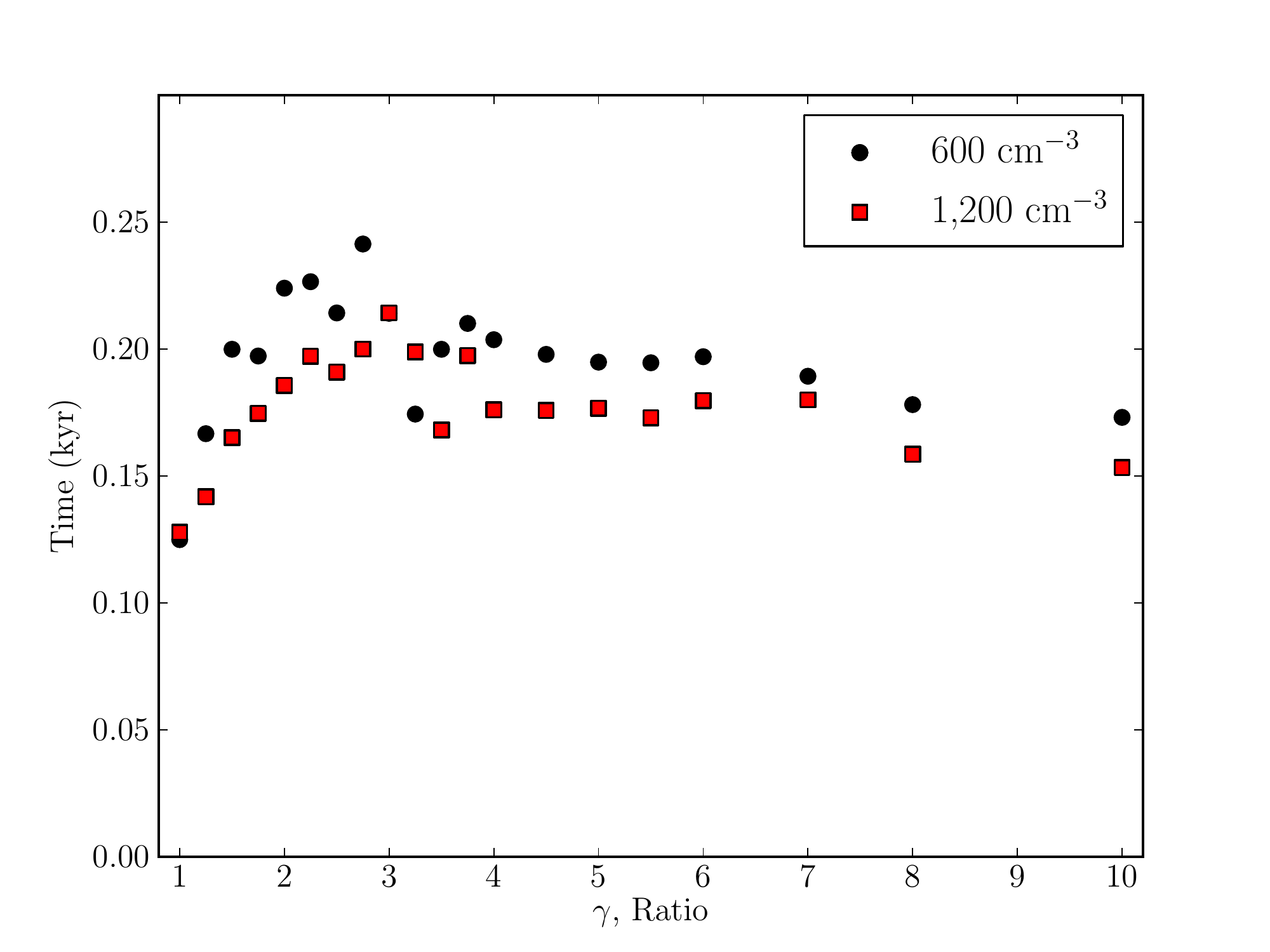}
\caption{The earliest core formation time for all clouds in the G1 and G2 series.}
\label{time_vs_ratio}
\end{figure}

Figure \ref{time_vs_ratio} shows the variation of the characteristic high density core formation time (when the highest density reaches $\approx 10^{13}$ \htwoden, as described in Section \ref{corefinder}), $t_{\mathrm{core}}$, over $\gamma$ for the two group clouds.  An overall picture is that the core formation time is shorter in each of the G2 clouds than in each corresponding G1 cloud of same $\gamma$, because the evolution of the G2 clouds is more shock dominated than that in the corresponding clouds in the G1 set.  Therefore less time is required to form a high density core. 
The variation in initial cloud density (600 \& 1,200 \htwoden) is small compared to the final densities (of order $10^{12}$ \htwoden), meaning that the increase in starting density alone for the G2 clouds relative to the G1 clouds is unlikely to account for the reduction in formation time. 

In both series, spherical clouds can be RDI shocked to form condensed cores in the shortest time of $\approx 0.125$ Myr. As $\gamma$ increases from 1, the core formation time increases. This is determined to be because, as the single focus splits towards two foci, collection of material, and subsequent gravitational collapse becomes slower. We take clouds in the G2 set as an example to look at the variation of $t_{\mathrm{core}}$ over $\gamma$. It is seen that $t_{\mathrm{core}}$ increases from 0.125 to 0.22 Myr in clouds of $ 1 < \gamma \le 3 $. For clouds with axial ratio  $ 3 < \gamma $,  they become increasingly elongated and the shocked gas has a decreasing distance to travel to collapse toward the foci, then the time needed for high density core formation decreases with $\gamma$. The variation of $t_{\mathrm{core}} $ vs $\gamma$ in G1 clouds is observed to follow a similar pattern to that in G2 clouds.

\subsection{The correlation with observation}
\begin{figure}
\center
\includegraphics[width=0.45\textwidth]{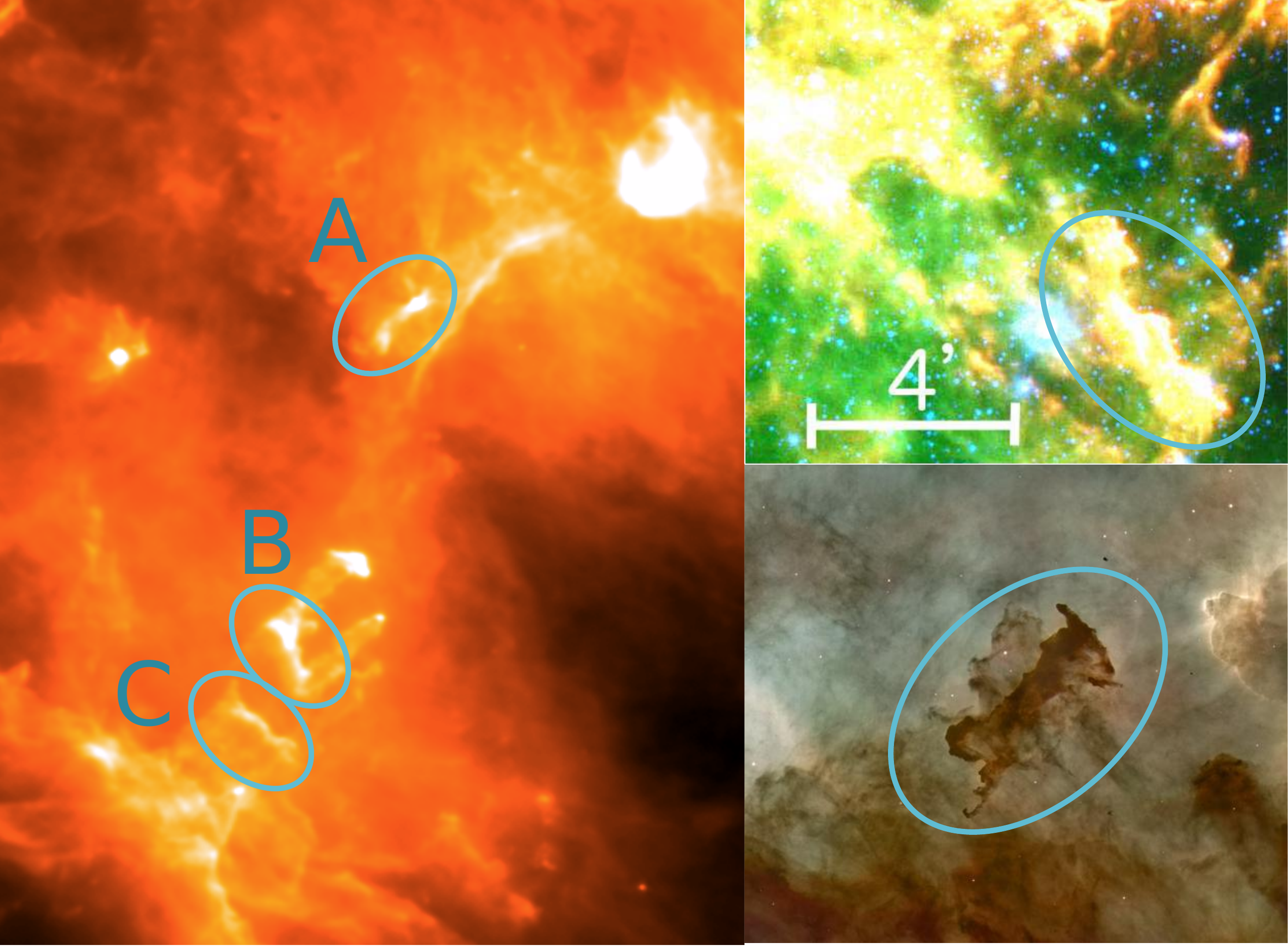}
\caption{Fragment-core structures taken from various sources. On the left is the segment of the Herschel image of M16 at 60 $\mu$m. In the upper right panel is a section of a Spitzer pseudo-color image of part of IC1848E \citep{ChauhanEtAl2011-1}. The lower right panel shows a segment of a fragment-core structure in a Hubble image of Carina Nebula in neutral hydrogen, taken from Hubble website.}
\label{fragments}
\end{figure}

Many of the fragment-core structures found at \HII\ boundaries have their linear axes perpendicular to the direction of the host star(s).  A few examples of such structures are presented in Figure \ref{fragments}. The morphology of these structures is very similar to that in the simulations we present in this paper. Object A in the left panel of Figure \ref{fragments} (a 60$\mu$m Herschel image of M16) is a typical linear structure with two condensed ends, whose morphological image is similar to the simulated structures from G2 clouds. Objects B and C in the same panel, as well as the other linear structures in the upper and lower right panels in Figure \ref{fragments}, have similar morphological structures to clouds A, B, C and several clouds in the G1 series. Therefore, it is reasonable to suggest that these fragment-core structures are the outcomes of the interplay between the EUV radiation from nearby stars and its initial prolate molecular cloud. 

However a quantitative comparison on the physical properties between simulation results and observations is not yet possible at this stage, due to lack of the detailed observational data. 

\subsection{Link to other modelling work}
The fragment-core structure found along \HII\ boundaries and the perfect \HII\ bubble structure \citep{WhitworthEtAl1994-1, DeharvengEtAl2009-1, DeharvengEtAl2012-1}  were taken as the result of the `Collect and Collapse' (C \&\ C) mechanism \citep{ElmegreenLada1977-1,DaleEtAl2007-1} in the previous theoretical modelling work. By setting  a star in the centre of a uniform spherical cloud, C \&\ C simulation can result in a perfect `Bubble'-like \HII\ region with a fragment-core inner boundary. 

Recently, \citet{WalchEtAl2012-1} performed SPH simulations based on RDI model, by replacing the uniform density spherical cloud used in C \&\ C model with a fractal molecular cloud.  Their simulations revealed the formation of a  similar \HII\ bubble structure with a wide spread network of fragment-core structure.  

Our simulations show that the fragment-core structure sporadically located along an \HII\  boundary could also be the consequence of RDI on a pre-existing uniform prolate cloud with its semi-major axis perpendicular to the ionising radiation flux.  

The RDI and C \&\ C mechanisms are equivalent in terms of the physical interaction process between ionisation radiation and a molecular cloud, but they are different in terms of the initial conditions of the molecular cloud used and the relative position of the star to the molecular cloud. A uniform spherical cloud with ionising star in its centre is used in C \&\ C model,  a fractal and spherical molecular cloud with stars at its centre is used in Walch's RDI model, and a pre-existing  prolate cloud with ionising stars at its one side is used in our RDI model.  The details resolved from different models could explain the variety in the structures of \HII\ regions observed.

\section{Conclusions}
Simulation results on three high mass prolate clouds reveal that a plane-parallel EUV radiation can trigger formation of distinctive fragment-core structure, in comparison with the formation of a high density spindle when no EUV radiation is present. 
    
Further investigation on both the high and low mass clouds finds the embedded cores can either spread over the final linear structure or accumulate around the two foci of the cloud, dependent on the initial conditions and radiation fluxes.   A dimensionless parameter of the EUV radiation flux penetration depth $d_{\mathrm{EUV}}$ can be used as an indicator to the evolutionary destiny of the clouds investigated.  In clouds of $d_{\mathrm{EUV}} \le 0.86 \%$, the collapse of a cloud is through foci convergency. The high density cores mainly locate around the two ends (two gravitational foci) of the linear structure with potential to form two well separate stars or two groups of stars.  In clouds of  $ 0.86 \% < d_{\mathrm{EUV}} \le 1.26 \%$, the mode of the cloud collapse is a mixture of foci and linear convergency.  The high density cores are found at one or two ends of the linear structure, while some cores with slightly lower centre density are also found between the two foci. In clouds of $d_{\mathrm{EUV}} > 1.26 \%$, the cloud collapses in the mode of linear convergency, when the high density cores spread over the whole linear structure with potential to form a chain of stars. 

Data analysis on the total core mass and core formation time in the two groups of low mass clouds (the G1 group with initial density of 600 \htwoden\ , G2 with that of 1200 \htwoden\ ) find that: i) the total core mass  $m_{\mathrm{tot}}$ in each of the G2 clouds is more than double that in each corresponding G1 cloud. ii) In clouds of same initial density,  $m_{\mathrm{tot}}$ decreases with $\gamma$ in clouds where there is only one or quasi-one gravitation centre, then a sharp increase in the cloud there are two well separated foci, finally decreases with with $\gamma$ again after $\gamma > 2.25$; ii)  The characteristic core formation time $t_{\mathrm{core}} $ is shorter in 95\% of the G2 clouds than that in the corresponding cloud in the G1 series. It increases with $\gamma$ when $\gamma \le 3$ , then becomes a quasi-constant at $\gamma > 3$ in both cloud groups; iii) the spherical cloud has the highest $m_{\mathrm{tot}}$ and shortest  $t_{\mathrm{core}}$ in both groups of clouds, which implies that EUV radiation triggered star formation in spherical cloud is most efficient.  

As the high density cores are the potential sites for future star formation, we can conclude that, for prolate clouds with their major-axis perpendicular to the same incident EUV radiation: (i) in clouds of the same axial ratio, EUV radiation triggered star formation would be more efficient in the cloud with higher initial density; (ii) in a group of clouds with same initial density, EUV radiation triggered star formation is more effective in clouds of intermediate axial ratio $ 1.75 \le \gamma < 3$. 

The sporadic core-fragment structures found in multiple \HII\ boundaries may be taken as the result of RDI in pre-existing prolate clouds, such as investigated here.
 
In our next paper, we will discuss the evolution of a prolate cloud inclined to the direction of the incident ionising radiation to address the mechanism for the formation of the BRCs with asymmetrical morphologies.   
 
\label{conclusion}

\bibliography{bib}
\bibliographystyle{mn2e}

\label{lastpage}

\end{document}